\documentclass{article}

\usepackage[preprint,nonatbib]{neurips_2025}
\usepackage{graphicx}
\usepackage{amsmath}

\usepackage[utf8]{inputenc} % allow utf-8 input
\usepackage[T1]{fontenc}    % use 8-bit T1 fonts
\usepackage{hyperref}       % hyperlinks
\usepackage{url}            % simple URL typesetting
\usepackage{booktabs}       % professional-quality tables
\usepackage{nicefrac}       % compact symbols for 1/2, etc.
\usepackage{microtype}      % microtypography
\usepackage[table,xcdraw]{xcolor}
\usepackage{amssymb}
\usepackage{wasysym}
\usepackage{utfsym}
\usepackage{adjustbox}
\usepackage{multirow}
\usepackage{enumitem}

\newcommand{\checkmarkgreen}{\textcolor[rgb]{0.1,0.6,0.2}{\usym{2714}}}

\newcommand{\checkmarkred}{\textcolor[rgb]{0.9,0,0}{\usym{2718}}}

\title{WavReward: Spoken Dialogue Models With Generalist Reward Evaluators}

\author{
Shengpeng Ji~$^{\spadesuit}$\thanks{Work done during internship at Alibaba Qwen Team.}
~~Tianle Liang
~~Yangzhuo Li
~~Jialong Zuo~$^{\spadesuit}$
\\
\textbf{
Minghui Fang~$^{\spadesuit}$
~~Jinzheng He~$^{\heartsuit}$
~~Yifu Chen~$^{\spadesuit}$ 
~~Zhengqing Liu
~~Ziyue Jiang~$^{\spadesuit}$
}
\\
\textbf{
Xize Cheng~$^{\spadesuit}$ 
~~Siqi Zheng
~~Jin Xu~$^\heartsuit$\thanks{Corresponding author.}
~~Junyang Lin~$^{\heartsuit}$ 
~~Zhou Zhao~$^\spadesuit$\footnotemark[2]
} \\ \\
$^\spadesuit$~Zhejiang University \& $^{\heartsuit}$~Alibaba Group \\
}

\begin{document}

\maketitle

\begin{abstract}
  End-to-end spoken dialogue models such as GPT-4o-audio have recently garnered significant attention in the speech domain. However, the evaluation of spoken dialogue models' conversational performance has largely been overlooked. This is primarily due to the intelligent chatbots convey a wealth of non-textual information which cannot be easily measured using text-based language models like ChatGPT. To address this gap, we propose \textbf{WavReward}, a reward feedback model based on audio language models that can \textbf{evaluate both the IQ and EQ of spoken dialogue systems with speech input}. Specifically, 1) based on audio language models, WavReward incorporates \textbf{the deep reasoning process and the nonlinear reward mechanism} for post-training. By \textbf{utilizing multi-sample feedback via the reinforcement learning} algorithm, we construct a specialized evaluator tailored to spoken dialogue models. 2) We introduce ChatReward-30K, a preference dataset used to train WavReward. ChatReward-30K includes both comprehension and generation aspects of spoken dialogue models. These scenarios span various tasks, such as text-based chats, nine acoustic attributes of instruction chats, and implicit chats. WavReward outperforms previous state-of-the-art evaluation models across multiple spoken dialogue scenarios, achieving a substantial improvement about Qwen2.5-Omni in objective accuracy from 53.4$\%$ to 91.5$\%$. In subjective A/B testing, WavReward also leads by a margin of 83$\%$. Comprehensive ablation studies confirm the necessity of each component of WavReward. \textbf{All data and code will be publicly at \url{https://github.com/jishengpeng/WavReward} after the paper is accepted}.
\end{abstract}

\section{Introduction}
Spoken dialogue models~\cite{ji2024wavchat} represent one of the most direct methods of human-computer interaction, evolving from traditional voice assistants such as Alexa, Siri, and Google Assistant to the latest intelligent dialogue systems, such as GPT-4o-audio\footnote{\url{https://openai.com/index/chatgpt-can-now-see-hear-and-speak/}}. Early spoken dialogue models~\cite{huang2024audiogpt, speechteam2024funaudiollm} were typically comprised of automatic speech recognition (ASR)~\cite{cao2012whisper, hsu2021hubert}, large language models (LLMs)~\cite{gpt4,touvron2023llama,bai2023qwen}, and text-to-speech (TTS)~\cite{ren2020fastspeech,kong2023vits2,jiang2023mega,jiang2024mega,shen2023naturalspeech,ji2024textrolspeech,ji2024controlspeech} components, which facilitated dialogue through a text-based cascading process that bridged speech input and output. To reduce latency and mitigate the cumulative errors of cascading systems, understand and generate non-textual paralinguistic information (e.g., emotion and sound) for real-time interaction, end-to-end spoken dialogue models~\cite{xie2024mini,fang2024llama,xie2024miniomni2opensourcegpt4ovision,wang2024freezeomnismartlowlatency,slamomni,minmo} such as GPT-4o-audio and Moshi~\cite{defossez2024moshi} have attracted considerable attention in both academic research and industry. By leveraging vast amounts of speech data~\cite{emilia, kahn2020libri} for multi-stage training, these end-to-end spoken dialogue models~\cite{minmo,defossez2024moshi, Qwen2.5-Omni} not only retain the intelligence quotient of text-based language models but also exhibit\textbf{ emotional quotient}, they are capable of handling diverse speech-related dialogue scenarios, such as role-playing, emotional dialogue, paralinguistic understanding, and paralinguistic controllable generation.

\begin{figure}[t]%%图,[htbp]是浮动格式
\centering
\includegraphics[height=11cm, width=13cm]{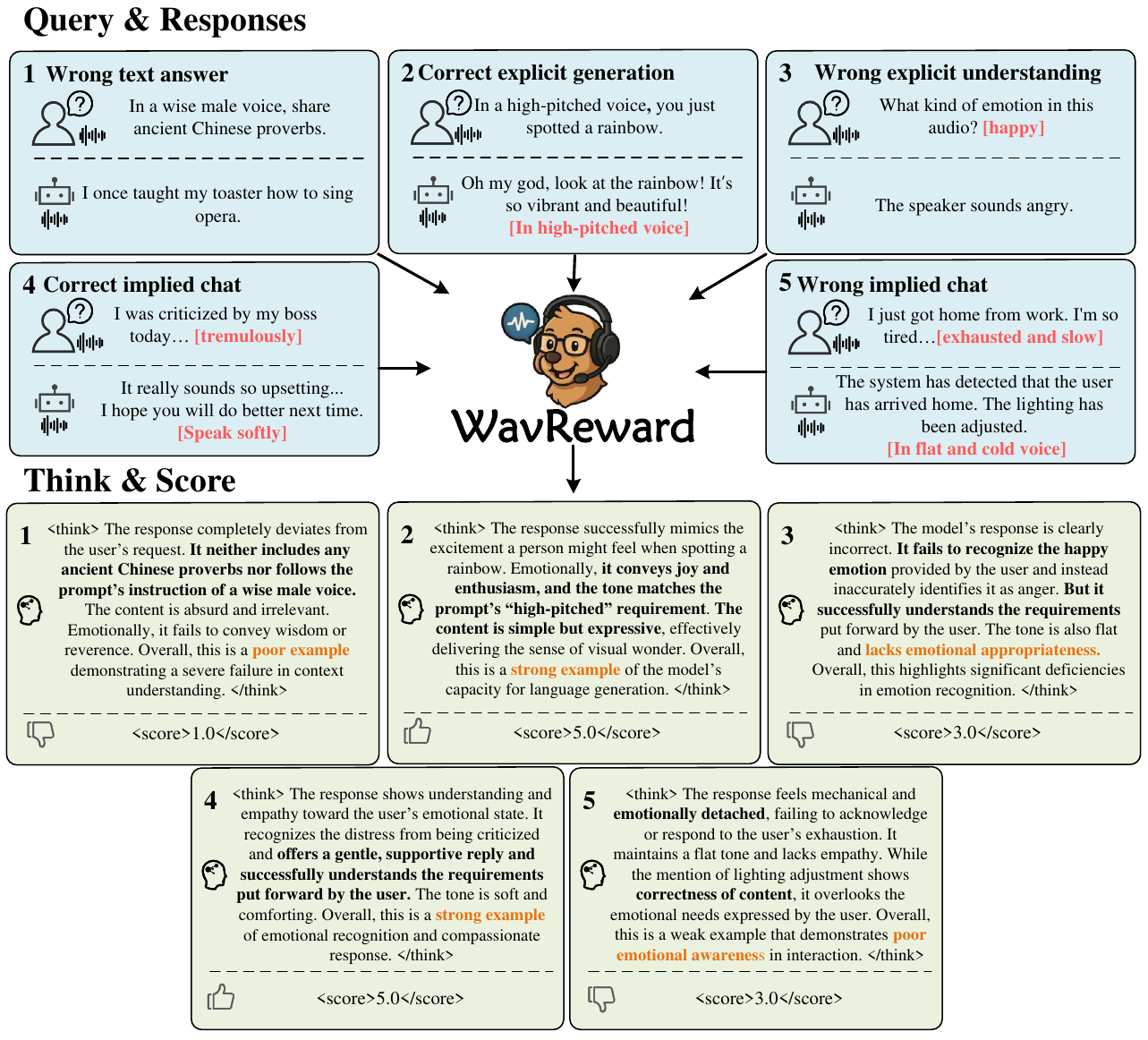}
\caption{WavReward can be applied to evaluate various dialogue scenarios, including both explicit instruction and implicit dialogues. It directly accepts speech-to-speech dialogue as input, evaluating the conversational coherence at both the textual and acoustic levels, and providing the final score.}
\label{figurejsp1}
\end{figure}

End-to-end spoken dialogue models~\cite{minmo, defossez2024moshi} have demonstrated remarkable conversational abilities, validating the potential of the speech modality in advancing toward Artificial General Intelligence. Thus, assessing the intelligence quotient and emotional quotient of end-to-end spoken dialogue models is a key challenge. This evaluation task involves three main challenges: 1) \textbf{the understanding and generation of substantial non-textual acoustic information (e.g., emotion, accent, pitch and sound) often present in dialogue scenarios}, which is currently not well-supported by any dedicated evaluation datasets of dialogue benchmarks~\cite{voicebench,sdeval,voxdialogue,yang2024air}. 2) \textbf{Dialogue is inherently multi-dimensional and multi-label.} For example, responses from spoken dialogue models may vary in speech rate either faster or slower, without a singular correct answer during casual conversations. 3) \textbf{Non-textual information in dialogue is often implicit}. For instance, when user return home late exhausted after work, an intelligent spoken dialogue model should be able to recognize the fatigue from user's voice and respond with a gentle, empathetic tone. Current benchmarks for evaluating spoken dialogue models, such as VoiceBench~\cite{voicebench}, AirBench~\cite{yang2024air}, VoxDialogue~\cite{voxdialogue} and SD-Eval~\cite{sdeval} primarily focus on the accuracy of textual information in dialogue, similar to using models like ChatGPT to assess the coherence of conversational text. Evaluation of non-textual information is limited to fixed tasks, such as emotion classification, gender recognition, and audio event detection, which assess the model’s \textbf{understanding of the acoustic information} in the dialogue.

To address the gap in evaluating end-to-end spoken dialogue models, we propose the WavReward model and the ChatReward-30K dataset. WavReward is a novel framework where audio language models~\cite{chu2024qwen2,Qwen2.5-Omni} (speech-to-text) can serve as evaluators for end-to-end spoken dialogue models~\cite{xie2024mini,defossez2024moshi}. As shown in Figure~\ref{figurejsp1}, WavReward can directly assess the capabilities of spoken dialogue models in both textual and non-textual acoustic dimensions. We demonstrate that fine-tuning audio language models with multiple examples via reinforcement learning~\cite{dpo,ppo,grpo,r1aqa} enables WavReward to provide reasonable scores across various scenarios. Furthermore, incorporating chain-of-thought reasoning~\cite{cot,audioreasoner,audiocot} into the evaluation process of audio language models significantly aids WavReward in generating more accurate scores. To augment the discriminative capability of WavReward across diverse dialogue contexts, WavReward includes the nonlinear reward mechanism and the positive-negative multi-sample sampling mechanism in the post-training reinforcement learning phase. Additionally, we construct the ChatReward-30K dataset to train WavReward and evaluate the performance of various evaluators~\cite{tang2023salmonn,chu2024qwen2}. ChatReward-30K not only contains standard text-centric dialogue examples but also incorporates diverse acoustic information\footnote{gender, age, language, accent, pitch, speed, volume, emotion and audio} from end-to-end dialogues. Each speech-to-speech dialogue sample in ChatReward-30K includes multiple responses to the same query. To our knowledge, this is the first dataset capable of comprehensively evaluating both the acoustic capabilities and the implicit conversational abilities of end-to-end spoken dialogue systems. Compared to the original audio language models and the supervised finetuned evaluators, WavReward significantly outperforms these baselines in both in-domain and out-domain scenarios. Furthermore, in human subjective A/B tests, WavReward outperforms direct inference with Qwen2.5-Omni~\cite{Qwen2.5-Omni} by the margin of 83$\%$. In summary, our contributions are as follows:

\begin{itemize}
    \item{WavReward is the first reward model specifically designed for end-to-end spoken dialogue models. It accepts \textbf{speech-to-speech dialogues} as input and provides corresponding scores for a wide range of dialogue scenarios. WavReward demonstrates that \textbf{audio language models can serve as effective evaluators for spoken dialogue models.}}
    \item{WavReward further enhances the evaluative capability through the reasoning-based assessment process, nonlinear reward feedback, and the positive-negative diverse sample sampling mechanism during the reinforcement learning post training.}
    \item{We introduce ChatReward-30K, the first dataset designed for training and evaluating audio feedback models. Compared to previous datasets, ChatReward-30K enables \textbf{comprehensive evaluation of both the acoustic information and implicit dialogue capabilities.}}

\end{itemize}

\section{Related work}

\subsection{Spoken Dialogue Models}
Spoken dialogue models refer to large language models~\cite{bai2023qwen,touvron2023llama} capable of engaging in conversations through both speech input and speech output. Traditional spoken dialogue models, such as AudioGPT~\cite{huang2024audiogpt} and FunAudioLLM~\cite{speechteam2024funaudiollm}, typically employ a three-stage cascading approach to facilitate dialogue. In this process, speech input is first transcribed into text using an automatic speech recognition model~\cite{cao2012whisper}. The transcribed text is then processed by a text-based LLM such as ChatGPT, to generate a textual response, which is subsequently converted back into speech using a text to speech model~\cite{du2024cosyvoice,cosyvoice2,ji2024mobilespeech}. However, these cascaded models often suffer from issues such as high latency, cumulative errors, and an inability to process non-textual acoustic information, which limits their effectiveness. Consequently, end-to-end spoken dialogue models~\cite{defossez2024moshi,zhang2023speechgpt,llamaomni2} have garnered significant attention in recent months. These models eliminate the need for transcription into text and directly process speech using either semantic~\cite{cao2012whisper,hsu2021hubert,du2024cosyvoice} or acoustic representations~\cite{encodec,ji2024wavtokenizer,ji2024language} for understanding and generation. For instance, LLaMA-Omni~\cite{fang2024llama} utilizes a Whisper encoder combined with an adapter to process speech, and generates corresponding Hubert tokens based on the LLM, which are then upsampled to produce speech. IntrinsicVoice~\cite{zhang2024intrinsicvoice} introduces GroupFormer to optimize the structure of Hubert token generation, while Mini-Omni1/2~\cite{xie2024mini,xie2024miniomni2opensourcegpt4ovision} employs a delay-pattern approach~\cite{musicgen} to directly generate the corresponding SNAC~\cite{siuzdak2024snac} acoustic tokens. Other similar end-to-end spoken dialogue models include SLAM-Omni~\cite{slamomni}, Freeze-Omni~\cite{wang2024freezeomnismartlowlatency}, VITA1.5~\cite{vita15}, OpenOmni~\cite{openomni}. Concurrently, numerous end-to-end spoken dialogue models such as GLM-4-Voice~\cite{glm4voice}, Moshi~\cite{defossez2024moshi}, Qwen2.5-Omni~\cite{Qwen2.5-Omni}, MinMo~\cite{minmo}, and Kimi-Audio~\cite{kimiaudio} have demonstrated significant intelligence quotient and emotional quotient emerging from large-scale speech training datasets. Although these spoken dialogue models exhibit strong conversational performance, \textbf{there remains a substantial gap in the assessment of both intelligence quotient and emotional quotient. In this paper, we present the first reward model WavReward specifically designed for the evaluation of spoken dialogue models}.

\subsection{Benchmark for Spoken Dialogue Models}

Early benchmarks related to spoken language, such as AudioBench~\cite{wang2024audiobench}, SUPERB~\cite{superb}, and MMAU~\cite{mmau}, primarily focus on evaluating fixed tasks such as emotion recognition, and are not well-suited for assessing a model's conversational abilities. With the rapid development of end-to-end spoken dialogue models~\cite{defossez2024moshi,minmo}, numerous new benchmarks have emerged to evaluate these spoken dialogue models. AirBench~\cite{yang2024air} leverages ChatGPT to evaluate the differences between generated text of speech-to-text dialogue models~\cite{chu2024qwen2} and ground truth text at the text level. SpokenWOZ~\cite{spokenwoz} transcribes the audio of the conversation into text via ASR models, and then uses metrics like BLEU to assess the performance of text-based language models. VoiceBench~\cite{voicebench} transcribes the dialogue audio from speech-to-speech dialogue models~\cite{defossez2024moshi,xie2024mini} into text and utilizes ChatGPT to evaluate the models' general knowledge and instruction-following ability. VoxDialogue~\cite{voxdialogue} and SD-Eval~\cite{sdeval} further focus on the ability of speech-to-text dialogue models~\cite{chu2023qwen,chu2024qwen2} to understand paralinguistic information, using BLEU and other text-based metrics in conjunction with ChatGPT to assess whether speech-to-text dialogue models~\cite{tang2023salmonn} can generate different textual responses based on varying acoustic information from different users.

However, \textbf{the aforementioned benchmarks still rely on transcribing audio into text for evaluation and cannot directly assess the acoustic coherence in speech-to-speech dialogues}. For example, when a user returns home tired after a long day, and the spoken dialogue model responds with a cheerful tone mocking the user, "mocking with a cheerful tone" cannot be directly evaluated by text-based models such as ChatGPT. \textbf{WavReward is the first evaluation model that accepts speech input and can directly assess the acoustic dialogue between the user and the spoken dialogue model. It can handle a diverse range of acoustic information, multi-label scenarios, and implicit dialogues scenarios, directly evaluating the realism of the acoustic interactions}. In addition, ChatReward-30K is the first comprehensive dataset \textbf{supporting the evaluation of paralinguistic understanding and generation, as well as implicit dialogue scenarios}.

% \begin{figure}[t]%%图,[htbp]是浮动格式
% \centering
% \includegraphics[height=9cm, width=14cm]{Wavreward_arc2.pdf}
% \caption{The overall structure of WavReward. WavReward directly accepts speech-to-speech dialogue audio for evaluation. The architecture is based on the audio language model and is trained using reinforcement learning on group samples. Additionally, WavReward incorporates the Chain-of-Thought reasoning process (the center of the diagram), along with positive and negative multi-sample sampling in the top-left corner, and the nonlinear reward mechanism in the top-right corner.}
% \label{figurejsp2}
% \end{figure}

\section{Method}

\subsection{WavReward}
\begin{figure}[t]%%图,[htbp]是浮动格式
\centering
\includegraphics[height=9cm, width=14cm]{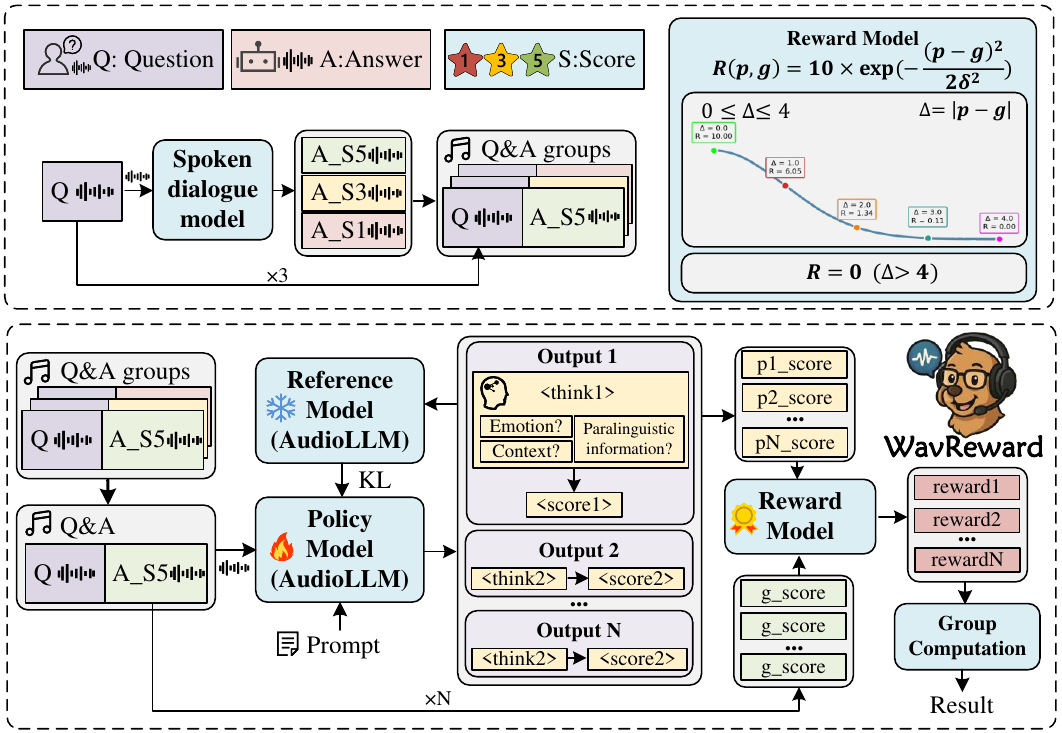}
\caption{The overall structure of WavReward. WavReward directly accepts speech-to-speech dialogue audio for evaluation. The architecture is based on the audio language model and is trained using reinforcement learning on group samples. Additionally, WavReward incorporates the Chain-of-Thought reasoning process (the center of the diagram), along with positive and negative multi-sample sampling in the top-left corner, and the nonlinear reward mechanism in the top-right corner.}
\label{figurejsp2}
\end{figure}

As shown in Figure~\ref{figurejsp2}, WavReward is an audio language model~\cite{Qwen2.5-Omni} that undergoes post-training fine-tuning through reinforcement learning~\cite{dpo,ppo,grpo,rl1}. In contrast to text-based large language models (LLMs) such as ChatGPT, audio language models~\cite{chu2024qwen2,Qwen2.5-Omni} can directly accept speech-to-speech dialogue as input, enabling a comprehensive evaluation of the coherence of both textual content and acoustic information in explicit and implicit dialogue scenarios. Similar to the conclusions drawn from reinforcement learning in text-based LLMs~\cite{grpo}, we find that fine-tuning with a small number of precise dialogue scoring samples via reinforcement learning significantly outperforms direct supervised fine-tuning. The relevant ablation results are presented in Table~\ref{tab:result}.

In the reward models of text-based LLMs, the primary task is to assess whether the content of question-answer pairs is reasonable, typically by sampling and providing feedback based on a single QA sample. However, in the speech dialogue, both the input and output contain abundant content and complex acoustic information. Single-sample QA feedback is insufficient for the reward model to effectively compare differences at various levels(content and acoustic). Therefore, \textbf{we design a multi-sample feedback mechanism in WavReward}, as shown in the top-left corner of Figure~\ref{figurejsp2}. For each dialogue scenario, we construct multiple answer-score pairs $\{a_{j},s_{j}\}$ at different levels for the given question $q$. The first level $s_{1}$ represents the content of answer that is deemed unreasonable and receives the lowest score. The second level $s_{2}$ evaluates the acoustic mismatch (e.g., when the user requests the spoken dialogue model to introduce U.S. history in a happy tone, but the model responds in an angry tone). Only when both the content and the acoustic information are correct will the dialogue receive the highest score $s_{3}$. Therefore, the input $x$ and target $y$ for WavReward during the training process are as follows:
\begin{equation}
    x=concat\left ( q,a_{j} \right ),\quad y=s_{j}, \quad  1 \le s_{j} \le 5, \quad j\in \left \{ 1,2,3 \right \} 
\label{equal: 1}
\end{equation}

Upon receiving the speech input $x$, WavReward initializes two policy models $W_{\theta }$ and 
$W_{\theta }^{'} $ with identical structures. Both $W_{\theta }$ and $W_{\theta }^{'} $ are speech-to-text audio language models~\cite{Qwen2.5-Omni}, where $W_{\theta }^{'} $ serves as the old policy model with frozen weights and the weights of the current training policy model $W_{\theta }$ remain updatable. Following the approach of DeepSeekMath~\cite{grpo}, we employ the Kullback-Leibler divergence loss to directly constrain the relationship between the reference policy model $W_{\theta }^{ref} $ and the current training policy model $W_{\theta }$ during the early stages of training. Notably, the KL divergence loss $\mathcal L_{KL}(W_{\theta },W_{\theta }^{ref})$ is not incorporated into the reward process of WavReward. The formulation is expressed as follows:
\begin{equation}
    \mathcal L_{KL}(W_{\theta },W_{\theta }^{'}) = \frac{W_{\theta }^{ref}(o_{i,t}|x,t_{prompt},o_{i,<t})}{W_{\theta }(o_{i,t}|x,t_{prompt},o_{i,<t})}-log\frac{W_{\theta }^{ref}(o_{i,t}|x,t_{prompt},o_{i,<t})}{W_{\theta }(o_{i,t}|x,t_{prompt},o_{i,<t})} - 1 
\end{equation}
where $t_{prompt}$ represents the text prompt for the policy model with specific examples provided in Appendix~\ref{appendix: prompt_think}, $t$ denotes the number of tokens, $o_{i}$ refers to the set of $N$ candidate outputs $\left \{ o_{1}, o_{2}, \dots ,o_{N} \right \} $ sampled by WavReward from the old policy model $W_{\theta }^{'}$ for each input $x$. It is important to note that each
$o_{i}$ in WavReward is not solely a score for evaluation. \textbf{We further incorporate a deep reasoning process by calculating the think format reward $R_{f}$(returning 5 or 0
based on compliance), which implicitly enables WavReward to analyze whether the responses $a_{i}$ of spoken dialogue models address the input question $q$ effectively from both content and acoustic perspectives,} and subsequently assign a final score $p$. WavReward computes the candidate rewards 
$\left \{ r_{1}, r_{2}, \dots ,r_{N} \right \} $
  for $N$ candidate outputs by comparing the $N$ candidate scores $\left \{ p_{1}, p_{2}, \dots ,p_{N} \right \} $ with the ground truth score 
$g$ using the accuracy reward $R_{a}$. Considering the discrepancy between the acoustic and content information in speech dialogues (the challenge of accurately perceiving acoustic information and providing responses with appropriate acoustic features as compared to content accuracy), \textbf{we design a nonlinear accuracy reward $R_{a}$, as illustrated in the upper-right corner of Figure \ref{figurejsp2}. When the difference between candidate score $p$ and ground score $g$ increases, the reward $R_{a}$ decreases exponentially, encouraging WavReward to provide higher accuracy rewards to spoken dialogue models that exhibit both cognitive intelligence quotient and emotional quotient.} The explicit formulation of 
$R_{a}$ is as follows:
\begin{equation}
    R_{a}(p,q) =
\begin{cases}
10 \cdot \exp\left( - \frac{(p - g)^2}{2\sigma^2} \right) & 0 \le \left | p - g \right | \le 4 \\
0 & \quad \left | p - g \right |> 4
\end{cases}
\label{equ:3}
\end{equation}

% \begin{equation}
%     R_{a}(p, q) =
%     \begin{cases}
%         10 \cdot \exp\left( - \frac{(p - g)^2}{2\sigma^2} \right) & \text{if } 0 \le \left | p - g \right | \le 4 \\
%         0 & \text{if } \left | p - g \right | > 4
%     \end{cases}
% \end{equation}
After obtaining $N$ candidate accuracy rewards 
$\left \{ r_{1}, r_{2}, \dots ,r_{N} \right \} $ through $R_{a}$, WavReward normalizes these accuracy rewards $r_{i}$ using the mean and standard deviation to derive the corresponding $A_{i}$:
\begin{equation}
A_i = \frac{r_i - \frac{1 }{N}\sum_{i=1}^{N} r_{i}  }{\sqrt{\frac{1 }{N}\sum_{i=1}^{N}(r_{i}-\frac{1 }{N}\sum_{i=1}^{N} r_{i})^{2} } }
\end{equation}
where $A_{i}$ represents the advantage of the candidate output score $p_{i}$ relative to other sampled output. Following~\cite{deepseekr1,deepseekv2,deepseekv3,grpo}, WavReward encourages the model to generate responses with higher advantages within the group $N$ by updating the policy model $W_{\theta}$ using the following objective $\mathcal{J}_{WavReward}(\theta)$, where $\epsilon$ and $\beta$ are hyper-parameters:

% \begin{equation}
% \footnotesize
% \begin{split}
% \mathcal{J}_{WavReward}&(\theta) = \mathbb{E} [x \sim P(X), \{o_i\}_{i=1}^{N} \sim W_{\theta}^{'}(O|x) ] \\
% & \frac{1}{N} \sum_{i=1}^{N} \frac{1}{|o_i|} \sum_{t=1}^{|o_i|}\left \{ min\left [ \frac{W_{\theta }(o_{i,t}|x,o_{i,<t})}{W_{\theta }^{'}(o_{i,t}|x,o_{i,<t})} A_{i,t}-clip\left ( \frac{W_{\theta }(o_{i,t}|x,o_{i,<t})}{W_{\theta }^{'}(o_{i,t}|x,o_{i,<t})} , 1-\epsilon ,1 + \epsilon \right )A_{i,t} \right ] \\ &-\beta \text{D}_{KL}[W_{\theta} || W_{\theta }^{ref}]  \right \} 
% \end{split}
% \end{equation}

\begin{equation}
\footnotesize
\begin{split}
\mathcal{J}_{WavReward}&(\theta) = \mathbb{E} [x \sim P(X), \{o_i\}_{i=1}^{N} \sim W_{\theta}^{'}(O|x) ] \\
& \frac{1}{N} \sum_{i=1}^{N} \frac{1}{|o_i|} \sum_{t=1}^{|o_i|} \Bigg\{ \min \left[ \frac{W_{\theta }(o_{i,t}|x,o_{i,<t})}{W_{\theta }^{'}(o_{i,t}|x,o_{i,<t})} A_{i,t} , \text{clip}\left( \frac{W_{\theta }(o_{i,t}|x,o_{i,<t})}{W_{\theta }^{'}(o_{i,t}|x,o_{i,<t})} , 1-\epsilon , 1 + \epsilon \right) A_{i,t} \right]  \\
& - \beta \text{D}_{KL}[W_{\theta} || W_{\theta }^{ref}] \Bigg\}
\end{split}
\end{equation}

\subsection{ChatReward-30K}

\subsubsection{The Overall of ChatReward-30K}
Given the absence of end-to-end dialogue datasets incorporating scores, we have developed and made available a dataset called ChatReward-30K, which contains spoken dialogue data across various scenarios along with corresponding scores. As shown in Table~\ref{tab:overview_chatreward}, ChatReward-30K demonstrates comprehensive coverage compared to existing evaluation datasets~\cite{voxdialogue,sdeval} for spoken dialogue models in the following key areas. \textbf{1) Evaluation from both content and acoustic dimensions.} Unlike previous datasets~\cite{voicebench}, ChatReward-30K evaluates dialogue performance from both content and acoustic perspectives, encompassing a wide range of paralinguistic features, including gender, age, language, accent, pitch, speed, volume, energy, emotion and audio. \textbf{2) Inclusion of both understanding and generation.} Previous datasets like Voxdialogue and SD-Eval primarily focus on the understanding component (speech-to-text) of spoken dialogue systems. In contrast, ChatReward-30K also evaluates the generation component, providing scenarios that assess how dialogue models generate speech in specific tones, such as speaking in the sad manner. \textbf{3) End-to-end implicit dialogue inclusion.} To further assess the emotional intelligence of spoken dialogue models, ChatReward-30K includes implicit dialogues across a variety of scenarios. For instance, it includes a scenario where a voice assistant offers gentle, empathetic comfort at a slow speech rate when the user is crying due to criticism from their boss. \textbf{4) Inclusion of both positive and negative examples.} To better train the WavReward model, as outlined in Equation~\ref{equal: 1}, ChatReward-30K features different dialogue responses for the same user scenario, providing both positive and negative examples to facilitate more effective model training. \textbf{5) Human expert scoring.} Each dialogue scenario in ChatReward-30K is accompanied by human expert ratings, ensuring that the scores reflect reasonable and well-founded assessments of dialogue quality.

\begin{table}[htbp]
    \centering
    \small
   \caption{Comparison of different evaluation dataset/benchmark for spoken dialogue models. \textbf{Dia.} refers to spoken \textbf{dialogue} and pure question-answering evaluation is not categorized as the dialogue (chat) task. \textbf{S2S.} denotes evaluation of \textbf{speech-to-speech} models. \textbf{Imp.} indicates \textbf{implicit} dialogues. \textbf{Neg. and Sco.} represent whether all positive and \textbf{negative} samples in the evaluation data are \textbf{scored}. As there is currently no dedicated dataset for training reward models, all datasets in this area are empty. Acoustic Information covers aspects like age, accent (acc.), gender (gen.), language (lan.), emotion (emo.), volume (vol.), speech rate (spe.), pitch (pit.) and audio (aud.). }
   % \wen{Please be consistent with how we refer to the three audio domains, esp. on using general sound or audio.}}
   \begin{adjustbox}{width=\textwidth}
    \begin{tabular}{c|ccccc|ccccccccc}
      \toprule
      \multirow{2}{*}{\textbf{Dataset/Benchmark}} & \multirow{2}{*}{\textbf{Dia.}} & \multirow{2}{*}{\textbf{S2S.}} & \multirow{2}{*}{\textbf{Neg.}} & \multirow{2}{*}{\textbf{Imp.}} & \multirow{2}{*}{\textbf{Sco.}}&  \multicolumn{9}{c}{\textbf{Acoustic Paralinguistic Information}} \\\cline{7-15}
       &  &  &  &  & &\textbf{age.} & \textbf{acc.}& \textbf{lan.}& \textbf{gen.}& \textbf{emo.}& \textbf{pit.}& \textbf{spe.} & \textbf{vol.}& \textbf{aud.}\\
      \midrule
       SUPERB& \checkmarkred & \checkmarkred & \checkmarkred& \checkmarkred & \checkmarkred & \checkmarkred & \checkmarkred& \checkmarkred& \checkmarkred& \checkmarkgreen& \checkmarkred& \checkmarkred& \checkmarkred& \checkmarkred \\
       MMAU & \checkmarkred & \checkmarkred & \checkmarkred & \checkmarkred &\checkmarkred& \checkmarkred& \checkmarkred& \checkmarkred& \checkmarkred& \checkmarkgreen& \checkmarkred& \checkmarkred& \checkmarkred& \checkmarkgreen \\
       AudioBench & \checkmarkred & \checkmarkred & \checkmarkred & \checkmarkred & \checkmarkred & \checkmarkred& \checkmarkgreen& \checkmarkred& \checkmarkgreen& \checkmarkgreen& \checkmarkred& \checkmarkred& \checkmarkred& \checkmarkgreen \\
       \midrule
       AirBench & \checkmarkgreen & \checkmarkred & \checkmarkred & \checkmarkred & \checkmarkred & \checkmarkred& \checkmarkred& \checkmarkred& \checkmarkgreen& \checkmarkgreen& \checkmarkred& \checkmarkred& \checkmarkred& \checkmarkgreen\\
       SD-Eval & \checkmarkgreen & \checkmarkred & \checkmarkred & \checkmarkred & \checkmarkred &\checkmarkgreen& \checkmarkgreen& \checkmarkred& \checkmarkgreen& \checkmarkgreen& \checkmarkred& \checkmarkred& \checkmarkred& \checkmarkgreen\\
      VoiceBench & \checkmarkgreen & \checkmarkgreen & \checkmarkred & \checkmarkred & \checkmarkred& \checkmarkred& \checkmarkred& \checkmarkred& \checkmarkred& \checkmarkred& \checkmarkred& \checkmarkred& \checkmarkred& \checkmarkred  \\
      VoxDialogue & \checkmarkgreen & \checkmarkred & \checkmarkred & \checkmarkgreen & \checkmarkred &\checkmarkgreen& \checkmarkgreen& \checkmarkgreen& \checkmarkgreen& \checkmarkgreen& \checkmarkgreen& \checkmarkgreen& \checkmarkgreen& \checkmarkgreen  \\
      \midrule
      ChatReward-30K & \checkmarkgreen & \checkmarkgreen & \checkmarkgreen & \checkmarkgreen & \checkmarkgreen & \checkmarkgreen& \checkmarkgreen& \checkmarkgreen& \checkmarkgreen& \checkmarkgreen& \checkmarkgreen& \checkmarkgreen& \checkmarkgreen& \checkmarkgreen \\
      \bottomrule
    \end{tabular}
    \end{adjustbox}
    \label{tab:overview_chatreward}
\end{table}

\subsubsection{Dataset Statistics}

ChatReward-30K consists of the total of 30000 samples, each dialogue sample represents the simulated user-chatbot interaction in the form of the speech-to-speech pair. Each dialogue is rated by human experts on a scale from 1 to 5, with the duration of each dialogue audio ranging from 5 to 35 seconds. ChatReward-30K is primarily divided into four components. 15$\%$ of the ChatReward-30K focuses on the textual aspects of the conversation. Based on the coherence of the dialogue content, another 25$\%$ of ChatReward-30K addresses the explicit understanding of user paralinguistic features such as recognizing when a child is interacting with the spoken dialogue model. The remaining 35$\%$ of ChatReward-30K pertains to the model’s generation ability of paralinguistic features such as adjusting the volume of the model's voice upon user request. The final 25$\%$ is unique to ChatReward-30K, representing implicit conversational scenarios such as the spoken dialogue model's ability to automatically detect the user’s emotional state and respond appropriately. Detailed examples can be found in Appendix~\ref{appendix: example_chatre}.

Following Equation~\ref{equal: 1}, each dialogue sample contains both positive and negative responses for the same user input. In terms of content, the dialogues in ChatReward are more aligned with natural and daily conversations rather than explicit QA pairs. As shown in Figure~\ref{figure_chatreward_data3} in Appendix~\ref{appendix_statistics}, the word cloud visualization of ChatReward-30K demonstrates a prevalence of natural spoken words, such as "can't" which is representative of daily spoken interactions. Concerning the acoustic attributes of the dialogues, most attribute categories in ChatReward-30K exhibit a relatively balanced distribution, as shown in Figure~\ref{figure_chatreward_data2} in Appendix~\ref{appendix_statistics}. Given the subtle emotional cues that humans can perceive in dialogue models, ChatReward-30K assigns particular emphasis to emotional attributes. Detailed information on each acoustic category is provided in Appendix~\ref{appendix: acoustic_in}. The ChatReward dataset is ultimately split into ChatReward-30K-train and ChatReward-30K-test sets with ratios of 85$\%$ and 15$\%$ respectively.

\subsubsection{Dataset Construction Process}

\textbf{Stage1: Dialogue Text Generation.} We begin by utilizing the GPT-4~\cite{gpt4} large language model to generate the text portion of the ChatReward-30K dataset through prompt engineering~\cite{promptprogramming}. To ensure the diversity of the dialogue content, we dynamically embed various topics, such as \textit{daily life, health management, education, entertainment, family relations, dietary culture, healthcare, shopping, internet usage, fitness, career development, and social interaction} during the text generation process. To generate explicit instruction-based dialogue data, we instruct the language model to generate dialogues that \textbf{contain various metalinguistic information}. For implicit dialogue data, we require the language model to annotate the generated conversation texts \textbf{with associated metalinguistic labels}. In alignment with Equation~\ref{equal: 1}, the ChatReward-30K dataset simulates dialogues between the same user and different model responses. This prompt template is as follows: \textit{ ModelA text must be relevant, high-quality content matching the required $< \text{label\_question}>$ (where $< \text{label\_question}>$ represents the expected metalinguistic label of the question); ModelB text must be relevant, high-quality content (similar quality to A), but using the $< \text{label\_b}>$ (where $< \text{label\_b}>$ represents a deliberately incorrect metalinguistic label used in the B-type response); ModelC text must be irrelevant and incorrect content (ensure that C is highly diverse and unique each time).} We then assign the score of 1 to dialogues with incorrect content, the score of 3 to dialogues with incorrect metalinguistic labels, and the score of 5 to dialogues with both correct content and metalinguistic labels. We observed that the prompt template significantly influences the quality and diversity of the generated dialogue. Therefore, we had human experts continuously adjust the prompt templates based on a small-scale dataset before further scaling up. Prompt programming templates can be found in Appendix~\ref{appendix: prompt}.

\textbf{Stage2: Dialogue Speech Generation.} In the generation process, we carefully tailor the most suitable SOTA TTS models for each attribute. We designed customized voice dialogue synthesis pipelines for each attribute to ensure the synthesized dialogue data accurately matches the corresponding attributes: \textbf{1) Accent, Pitch, and Emotion}: we utilize GPT-4o-mini-TTS to generate conditionally based speech by adjusting stylistic instructions. This tool focuses on speech techniques such as tongue-twisting, pauses, breathing, and whispering to accurately produce accents and emotions. Based on ten built-in speaker timbre, the model is instructed to synthesize speech using the following command format: Repeat this sentence with the <emotion>/<accent>/<pitch> of <example>. \textbf{2) Age:} we randomly selected 1000 speaker~\cite{agedata} samples from four age groups as reference voices. To minimize textual content discrepancies across different cloned voices, we selected cloned samples with different tones but identical dialogue content for four distinct age groups and used Step-Audio-TTS-3B~\cite{stepaudio} for voice cloning. \textbf{3) Speed, Volume, Gender, and Language:} we use CosyVoice2~\cite{cosyvoice2} to synthesize speech with specified voice characteristics. The volume and speech rate are adjusted using correlation coefficients to achieve the desired attributes. \textbf{4) Audio:} we combine instruct speech clips with audio clips together. Specifically, we selected 39 categories from the AudioCaps~\cite{kim2019audiocaps} dataset (500 samples per category) that include various audio events. Given that current dialogue models generally do not support music-related conversations, while audio events such as coughing, laughter, and crying are integral components of everyday interactions, our acoustic information category includes audio but excludes music. After synthesizing all the speech segments, we concatenate the simulated user speech segments with the simulated model response speech segments, ensuring a 1-second silence gap between them.

\textbf{Step3: Data Filtering and Scoring.} We used the Whisper-Large-V3~\cite{whisperopenai} model to filter out all sentences with the WER greater than 5$\%$. Given the large volume of emotional speech and the ambiguity in category boundaries, we utilized the Emotion2Vec~\cite{ma2023emotion2vec} model to filter out audio with inaccurate emotional labels and removed synthetic speech with scores below 0.5. To further improve the quality of the ChatReward-30K dataset, we invited five human experts to manually verify and adjust the text, speech, and scoring results of the dataset.

\section{Experiments}

\subsection{Experiment setup}

\textbf{Datasets.} Since there is currently no dataset available for training and evaluating evaluators for spoken dialogue models~\cite{xie2024mini}, we use ChatReward-30K-train as the training set for WavReward, with the ChatReward-30K-test subset reserved for testing. We evaluate the models using the ChatReward-30K-test (4000 samples) across three aspects: \textbf{content, explicit paralinguistic understanding and generation (with 9 distinct paralinguistic features), and implicit dialogue}. Additionally, we record 120 real human-machine dialogues between users and LLaMA-Omni~\cite{fang2024llama} (overall biased negative samples) and Kimi-Audio~\cite{kimiaudio} (overall biased positive samples), named the RealDialogue \textbf{to compare the performance of different evaluators in more realistic out-of-domain settings}. In the RealDialogue dataset, we observed that certain dialogues have extended durations, and there are instances of poor audio quality, such as distorted electronic sounds. \textbf{These factors present a more rigorous challenge for evaluating the model's performance in unseen, real-world scenarios.}\\
\textbf{Baselines.} Similar to using ChatGPT for assessing the coherence of text-based dialogues, we employ various audio language models~\cite{chu2024qwen2} (speech-to-text) as baseline evaluators to score speech-to-speech dialogues. The specific audio language models include Qwen-Audio~\cite{chu2023qwen}, SALMONN~\cite{tang2023salmonn}, Audio Flamingo2~\cite{audioflamingo2}, Qwen2-Audio~\cite{chu2024qwen2}, Qwen2.5-Omni~\cite{Qwen2.5-Omni} and GPT-4o-audio. Furthermore, we enhance two new versions by fine-tuning Qwen2.5-Omni using both full-parameter and LoRA~\cite{hu2022lora} fine-tuning methods on the WavReward-30K-train dataset. We utilized the official modelscopes fine-tuning framework ms-swift~\cite{swift} for this process. To ensure a fair comparison, all supervised fine-tuned models were trained for the same steps. \\
\textbf{Training details.} WavReward is trained using 8 GPUs (due to company policy, the specific details are confidential.), each running a batch size of 1, with gradient accumulation performed every 2 steps. The model is trained for 3500 steps with a learning rate of $1 \times 10^{-6}$ and a temperature of 1.0. The maximum number of cot tokens is set to 5120, and the weight coefficient for the KL loss is set to 0.01. The model architecture of WavReward is based on the sota open-source audio language model Qwen2.5-Omni-7B-Think~\cite{Qwen2.5-Omni} with the identical parameters. All parameters of WavReward are updated during the training process. \\
\textbf{Metric.} For evaluation on the ChatReward-30K-test and RealDialogue, we use accuracy to measure the difference between the predicted scores and the ground truth (GT) scores. On the RealDialogue-testset, we conduct subjective A/B testing via crowdsourcing, where 5 human experts are required to select the optimal score between the different scores given by WavReward and various baseline evaluators in the same real dialogue.

\subsection{Main Results}

\begin{table}[htbp]
    \centering
    \small
   \caption{The accuracy of scoring by WavReward and various baselines on the ChatReward-test and RealDialogue datasets is evaluated. Specifically, the ChatReward-test dataset is assessed across three main dimensions: content scoring, acoustic instruction dialogue scoring (which includes both understanding and generation), and implicit dialogue scoring. The acoustic information which are categorized as follows: age, accent (acc.), gender (gen.), language (lan.), emotion (emo.), volume (vol.), speech rate (spe.), pitch (pit.), and audio (aud.).}
 
   \begin{adjustbox}{width=\textwidth}
    \begin{tabular}{c|ccccccccccc|c}
      \toprule
      \multirow{2}{*}{\textbf{Model}} & \multirow{2}{*}{\textbf{Content}} &  \multicolumn{9}{c}{\textbf{Acoustic Instruction}}& \multirow{2}{*}{\textbf{Implicit}} & \multirow{2}{*}{\textbf{RealDialogue}} \\ \cline{3-11}
       &    &\textbf{age.} & \textbf{acc.}& \textbf{lan.}& \textbf{gen.}& \textbf{emo.}& \textbf{pit.}& \textbf{spe.} & \textbf{vol.}& \textbf{aud.}&&\\
      \midrule
      \multicolumn{13}{l}{\cellcolor{black!10} \textbf{\textit{I. Baseline audio langauge models direct inference with prompt}}} \\
      Qwen2-Audio & 24.7 & 32.4 & 24.5 & 36.8& 27.6& 33.7& 32.3& 40.5& 41.4& 50.8  & 28.9 & 42.5 \\
      Qwen-Audio & 43.4 & 35.5& 23.0& 33.6& 14.9& 34.2& 27.7& 35.4& 39.0& 32.2& 35.2 & 40.8 \\
      SALMONN & 13.5 & 33.9& 34.8& 28.4& 33.9& 36.4& 25.4& 30.3& 28.0& 51.3 & 20.3 & 19.2 \\
      Audio Flamingo2 & 22.7 & 20.8& 25.4& 16.8& 18.8& 18.6& 17.8& 21.5& 21.9& 20.6  & 22.6 & 21.6 \\
      GPT-4o-audio & 69.4 & 92.0 & 57.7& 100& 82.1 & 58.5& 88.7& 94.5& 88.1& 83.3  & 53.6 & 57.6 \\
      
      % Gemini-2.5 Pro & - & -& -& -& -& -& -& -& -& -  & - & - \\
      
      Qwen2.5-Omni & 54.6 & 56.3& 50.0& 48.4& 54.1& 57.8& 66.1& 34.1& 53.6& 64.9  & 48.5 & 51.7 \\
      \midrule
      \multicolumn{13}{l}{\cellcolor{black!10} \textbf{\textit{II. Baseline audio langauge models after supervised fine-tuning}}} \\
      Qwen2.5-Omni w/ Full-param tuning & 67.1 & 81.9 & 69.6& 98.9 & 83.8 & 66.5 & 76.1 & 84.8 & 86.6 & 86.7 & 55.8 & 58.3 \\
      Qwen2.5-Omni w/ LoRA & 63.8 & 81.2& 43.6& 100& 82.1 & 49.1& 74.1 & 83.6& 85.1& 85.7  & 54.2 & 56.7 \\
      \midrule
      \multicolumn{13}{l}{\cellcolor{black!10} \textbf{\textit{III. Different ablation versions of WavReward}}} \\
      WavReward w/o cot think & 84.2 & 80.3 & 77.9& 98.9 & 86.9& 85.4& 80.7& 86.0& 90.2& 88.6 & 61.4 & 59.1\\
      WavReward w/o multi samples & 85.3 & 85.7 & 69.6& 98.9& 88.6& 90.0& 82.1& 88.6& 85.4& 90.2 & 61.9 & 72.5\\
      WavReward w/o nonlinear reward& 88.6 & 92.2& 80.9& 100& 89.8& 90.5& 83.8& 87.3& 92.7& 85.7 & 66.6 & 70.8\\

      \midrule

      \textbf{WavReward (ours)} & \textbf{90.8} & \textbf{96.9}& \textbf{87.7}& \textbf{100}& \textbf{95.5}& \textbf{97.5}& \textbf{89.1}& \textbf{91.1}& \textbf{97.6}& \textbf{97.0}  & \textbf{74.3} & \textbf{80.8}\\
      \bottomrule
    \end{tabular}
    \end{adjustbox}
    \label{tab:result}
\end{table}

We evaluated the generation and ground-truth score accuracy of WavReward and Baseline models on the ChatReward-30K-test as well as the real out-of-domain RealDialogue dataset. The Baseline is divided into two categories: one consists of direct inference from audio language models using text prompt templates consistent with WavReward, and the other is the evaluator fine-tuned using the ChatReward-30K-train. The specific experimental results are presented in Table~\ref{tab:result}. We can draw the following conclusions: \textbf{1) WavReward significantly outperforms the best audio language models GPT-4o-audio on all metrics}. It achieved improvements of 21.4, 20.7, and 39.0 points in the content scoring, implicit dialogue scoring, and emotion-instructed dialogue scoring, respectively. Furthermore, it outperformed the direct inference model Qwen2.5-Omni by an average factor of two. This indicates that audio language models when optimized using reinforcement learning, can effectively serve as evaluators for spoken dialogue models. Moreover, RL significantly improves performance compared to direct inference. \textbf{2) We found that the RL-based WavReward surpassed the LoRA fine-tuned Qwen2.5-Omni}. This may be due to the direct scoring approach of supervised fine-tuning, which is overly simplistic and struggles to capture the complex scoring logic needed for various scenarios.  \textbf{3) We observed a substantial performance gap between different audio language models during direct inference}, highlighting the need for future work to develop and open-source more robust foundational audio language models. \textbf{4) We found that WavReward's accuracy in scoring accent-based sub-languages and implicit dialogues was lower than that for other scenarios.} This may be attributed to the less accurate accent data in the ChatReward-30K dataset compared to other acoustic information, and the inherent difficulty in evaluating implicit dialogues. There remains room for improvement in determining what constitutes a reasonable emotional response in spoken dialogue models.
\textbf{5) On the RealDialogue dataset, WavReward achieved a score accuracy of 80$\%$,} indicating that it exhibits strong robustness and can provide reliable evaluations in real-world, complex scenarios.

\subsection{A/B Test on RealDialogue}
Given that evaluating the responses of end-to-end spoken dialogue models in implicit dialogue settings constitutes a multi-dimensional, multi-label scenario, where there is no single ground truth  label, and the responses of dialogue models should ideally align with human subjective preferences, we have incorporated a subjective A/B testing approach. Specifically, five human experts were tasked with evaluating data from RealDialogue and determining which of two distinct discriminators provided the most reasonable assessment. To ensure the validity of the subjective criteria, experts were also asked to provide justification for their choices. We conducted pairwise comparisons between three baseline: Qwen2.5-Omni w/ direct inference, GPT-4o-audio w/ direct inference, and Qwen2.5-Omni w/ LoRA. The objective was to compare the different scoring outcomes of WavReward and the baseline model on the same test set from RealDialogue, with the results presented in Table~\ref{tab:ab_test}. Our findings indicate that WavReward outperformed Qwen2.5-Omni w/ direct inference in the subjective A/B test by a margin of 83$\%$, and also achieved a 77$\%$ success rate when compared to GPT-4o-audio w/ direct inference. These results suggest that WavReward's scoring is more closely aligned with human subjective preferences, demonstrating superior performance across a wide range of real-world dialogue scenarios.

\begin{table}[htbp]
    \centering
    \small
   \caption{Subjective A/B testing of the scoring reasonableness of WavReward and different evaluators on the RealDialogue dataset.}
   \begin{adjustbox}{width=0.65\textwidth}
    \begin{tabular}{c|cc}
      \toprule
      Models & WavReward Win $\uparrow $ & WavReward Lose $\downarrow $  \\
      \midrule
      Qwen2.5-Omni w/ direct inference & 83 & 17 \\
      Qwen2.5-Omni w/ LoRA & 79 &  21 \\
      GPT-4o-audio w/ direct inference & 77 &  23 \\
      \bottomrule
    \end{tabular}
    \end{adjustbox}
    \label{tab:ab_test}
\end{table}

\subsection{Ablation experiments}
\textbf{w/o cot think.} We removed the chain-of-thought (CoT) reasoning from WavReward and referred to this version as WavReward w/o CoT think. Specifically, WavReward in this configuration directly generates scores without the additional CoT-based format reward (loss) and the corresponding $t_{prompt}$ was also modified as shown in Appendix~\ref{appendix: prompt_think}. All other training and model parameters remain unchanged. As shown in Table~\ref{tab:result}, we found that CoT reasoning improved accuracy by approximately 10$\%$ across all evaluation categories. In out-of-domain scenarios, the improvement was as high as 21.7$\%$. This suggests that reasoning capabilities are beneficial for the evaluator model.

\textbf{w/o nolinear reward.} We replaced the reward function in Equation~\ref{equ:3} with a classic linear 0/1 reward function~\cite{deepseekv2,deepseekv3}. Specifically, the WavReward w/o nolinear reward version receives the reward of 5 when the generated score matches the ground truth score, and no reward is given when there is a mismatch during training. All other training and model parameters are consistent with the previous configuration. By comparing the versions of WavReward and WavReward w/o nonlinear in Table~\ref{tab:result}, we observe that the non-linear reward function aids WavReward in learning the differences in various levels of information in speech. For instance, when there is a large discrepancy between the GT score and the predicted score (e.g., a high emotional intelligence response receives a low score of 1 from WavReward), a substantial penalty is applied which helps the model correct such errors.

\textbf{w/o multi samples.} In classical reinforcement learning algorithms~\cite{ppo}, single-sample sampling can be used to calculate rewards based on the difference between the GT score and the generated score. In the WavReward w/o multi-samples version, for each question only one randomly selected answer is used for evaluation. All other training and model parameters remain unchanged. We found that performance dropped when multi-sample evaluation was removed. This decline can be attributed to the loss of the ability to simulate a range of reasonable and unreasonable responses to the same question, which assists WavReward in distinguishing between different scoring criteria and their variations.

\section{Conclusion}
In this work, we present WavReward, the first evaluation framework capable of supporting speech-to-speech input and providing comprehensive assessments of spoken dialogue models at both the text and acoustic levels. WavReward leverages reinforcement learning to turn audio language models into evaluatorsn and incorporate the chain-of-thought reasoning process, nonlinear rewards, and both positive and negative sample feedback to enhance the validity of the evaluation. In a variety of in-domain and out-of-domain explicit and implicit evaluation scenarios, WavReward significantly outperforms previous state-of-the-art evaluators. Furthermore, in human subjective A/B tests, it shows a substantial lead with the 83$\%$ improvement. In the future, we aim to scale up audio language models (e.g., 7B-70B) to further enhance WavReward's capabilities.

\bibliographystyle{unsrt}
\bibliography{neurips}

\begin{thebibliography}{10}

\bibitem{ji2024wavchat}
Shengpeng Ji, Yifu Chen, Minghui Fang, Jialong Zuo, Jingyu Lu, Hanting Wang, Ziyue Jiang, Long Zhou, Shujie Liu, Xize Cheng, et~al.
\newblock Wavchat: A survey of spoken dialogue models.
\newblock {\em arXiv preprint arXiv:2411.13577}, 2024.

\bibitem{huang2024audiogpt}
Rongjie Huang, Mingze Li, Dongchao Yang, Jiatong Shi, Xuankai Chang, Zhenhui Ye, Yuning Wu, Zhiqing Hong, Jiawei Huang, Jinglin Liu, et~al.
\newblock Audiogpt: Understanding and generating speech, music, sound, and talking head.
\newblock In {\em Proceedings of the AAAI Conference on Artificial Intelligence}, volume~38, pages 23802--23804, 2024.

\bibitem{speechteam2024funaudiollm}
Tongyi SpeechTeam.
\newblock Funaudiollm: Voice understanding and generation foundation models for natural interaction between humans and llms.
\newblock {\em arXiv preprint arXiv:2407.04051}, 2024.

\bibitem{cao2012whisper}
Nan Cao, Yu-Ru Lin, Xiaohua Sun, David Lazer, Shixia Liu, and Huamin Qu.
\newblock Whisper: Tracing the spatiotemporal process of information diffusion in real time.
\newblock {\em IEEE transactions on visualization and computer graphics}, 18(12):2649--2658, 2012.

\bibitem{hsu2021hubert}
Wei-Ning Hsu, Benjamin Bolte, Yao-Hung~Hubert Tsai, Kushal Lakhotia, Ruslan Salakhutdinov, and Abdelrahman Mohamed.
\newblock Hubert: Self-supervised speech representation learning by masked prediction of hidden units.
\newblock {\em IEEE/ACM transactions on audio, speech, and language processing}, 29:3451--3460, 2021.

\bibitem{gpt4}
Josh Achiam, Steven Adler, Sandhini Agarwal, Lama Ahmad, Ilge Akkaya, Florencia~Leoni Aleman, Diogo Almeida, Janko Altenschmidt, Sam Altman, Shyamal Anadkat, et~al.
\newblock Gpt-4 technical report.
\newblock {\em arXiv preprint arXiv:2303.08774}, 2023.

\bibitem{touvron2023llama}
Hugo Touvron, Thibaut Lavril, Gautier Izacard, Xavier Martinet, Marie-Anne Lachaux, Timoth{\'e}e Lacroix, Baptiste Rozi{\`e}re, Naman Goyal, Eric Hambro, Faisal Azhar, et~al.
\newblock Llama: Open and efficient foundation language models.
\newblock {\em arXiv preprint arXiv:2302.13971}, 2023.

\bibitem{bai2023qwen}
Jinze Bai, Shuai Bai, Yunfei Chu, Zeyu Cui, Kai Dang, Xiaodong Deng, Yang Fan, Wenbin Ge, Yu~Han, Fei Huang, et~al.
\newblock Qwen technical report.
\newblock {\em arXiv preprint arXiv:2309.16609}, 2023.

\bibitem{ren2020fastspeech}
Yi~Ren, Chenxu Hu, Xu~Tan, Tao Qin, Sheng Zhao, Zhou Zhao, and Tie-Yan Liu.
\newblock Fastspeech 2: Fast and high-quality end-to-end text to speech.
\newblock {\em arXiv preprint arXiv:2006.04558}, 2020.

\bibitem{kong2023vits2}
Jungil Kong, Jihoon Park, Beomjeong Kim, Jeongmin Kim, Dohee Kong, and Sangjin Kim.
\newblock Vits2: Improving quality and efficiency of single-stage text-to-speech with adversarial learning and architecture design.
\newblock {\em arXiv preprint arXiv:2307.16430}, 2023.

\bibitem{jiang2023mega}
Ziyue Jiang, Yi~Ren, Zhenhui Ye, Jinglin Liu, Chen Zhang, Qian Yang, Shengpeng Ji, Rongjie Huang, Chunfeng Wang, Xiang Yin, et~al.
\newblock Mega-tts: Zero-shot text-to-speech at scale with intrinsic inductive bias.
\newblock {\em arXiv preprint arXiv:2306.03509}, 2023.

\bibitem{jiang2024mega}
Ziyue Jiang, Jinglin Liu, Yi~Ren, Jinzheng He, Zhenhui Ye, Shengpeng Ji, Qian Yang, Chen Zhang, Pengfei Wei, Chunfeng Wang, et~al.
\newblock Mega-tts 2: Boosting prompting mechanisms for zero-shot speech synthesis.
\newblock In {\em The Twelfth International Conference on Learning Representations}, 2024.

\bibitem{shen2023naturalspeech}
Kai Shen, Zeqian Ju, Xu~Tan, Yanqing Liu, Yichong Leng, Lei He, Tao Qin, Sheng Zhao, and Jiang Bian.
\newblock Naturalspeech 2: Latent diffusion models are natural and zero-shot speech and singing synthesizers.
\newblock {\em arXiv preprint arXiv:2304.09116}, 2023.

\bibitem{ji2024textrolspeech}
Shengpeng Ji, Jialong Zuo, Minghui Fang, Ziyue Jiang, Feiyang Chen, Xinyu Duan, Baoxing Huai, and Zhou Zhao.
\newblock Textrolspeech: A text style control speech corpus with codec language text-to-speech models.
\newblock In {\em ICASSP 2024-2024 IEEE International Conference on Acoustics, Speech and Signal Processing (ICASSP)}, pages 10301--10305. IEEE, 2024.

\bibitem{ji2024controlspeech}
Shengpeng Ji, Jialong Zuo, Minghui Fang, Siqi Zheng, Qian Chen, Wen Wang, Ziyue Jiang, Hai Huang, Xize Cheng, Rongjie Huang, et~al.
\newblock Controlspeech: Towards simultaneous zero-shot speaker cloning and zero-shot language style control with decoupled codec.
\newblock {\em arXiv preprint arXiv:2406.01205}, 2024.

\bibitem{xie2024mini}
Zhifei Xie and Changqiao Wu.
\newblock Mini-omni: Language models can hear, talk while thinking in streaming.
\newblock {\em arXiv preprint arXiv:2408.16725}, 2024.

\bibitem{fang2024llama}
Qingkai Fang, Shoutao Guo, Yan Zhou, Zhengrui Ma, Shaolei Zhang, and Yang Feng.
\newblock Llama-omni: Seamless speech interaction with large language models.
\newblock {\em arXiv preprint arXiv:2409.06666}, 2024.

\bibitem{xie2024miniomni2opensourcegpt4ovision}
Zhifei Xie and Changqiao Wu.
\newblock Mini-omni2: Towards open-source gpt-4o with vision, speech and duplex capabilities, 2024.

\bibitem{wang2024freezeomnismartlowlatency}
Xiong Wang, Yangze Li, Chaoyou Fu, Lei Xie, Ke~Li, Xing Sun, and Long Ma.
\newblock Freeze-omni: A smart and low latency speech-to-speech dialogue model with frozen llm, 2024.

\bibitem{slamomni}
Wenxi Chen, Ziyang Ma, Ruiqi Yan, Yuzhe Liang, Xiquan Li, Ruiyang Xu, Zhikang Niu, Yanqiao Zhu, Yifan Yang, Zhanxun Liu, et~al.
\newblock Slam-omni: Timbre-controllable voice interaction system with single-stage training.
\newblock {\em arXiv preprint arXiv:2412.15649}, 2024.

\bibitem{minmo}
Qian Chen, Yafeng Chen, Yanni Chen, Mengzhe Chen, Yingda Chen, Chong Deng, Zhihao Du, Ruize Gao, Changfeng Gao, Zhifu Gao, et~al.
\newblock Minmo: A multimodal large language model for seamless voice interaction.
\newblock {\em arXiv preprint arXiv:2501.06282}, 2025.

\bibitem{defossez2024moshi}
Alexandre D{\'e}fossez, Laurent Mazar{\'e}, Manu Orsini, Am{\'e}lie Royer, Patrick P{\'e}rez, Herv{\'e} J{\'e}gou, Edouard Grave, and Neil Zeghidour.
\newblock Moshi: a speech-text foundation model for real-time dialogue.
\newblock {\em arXiv preprint arXiv:2410.00037}, 2024.

\bibitem{emilia}
Haorui He, Zengqiang Shang, Chaoren Wang, Xuyuan Li, Yicheng Gu, Hua Hua, Liwei Liu, Chen Yang, Jiaqi Li, Peiyang Shi, et~al.
\newblock Emilia: A large-scale, extensive, multilingual, and diverse dataset for speech generation.
\newblock {\em arXiv preprint arXiv:2501.15907}, 2025.

\bibitem{kahn2020libri}
Jacob Kahn, Morgane Riviere, Weiyi Zheng, Evgeny Kharitonov, Qiantong Xu, Pierre-Emmanuel Mazar{\'e}, Julien Karadayi, Vitaliy Liptchinsky, Ronan Collobert, Christian Fuegen, et~al.
\newblock Libri-light: A benchmark for asr with limited or no supervision.
\newblock In {\em ICASSP 2020-2020 IEEE International Conference on Acoustics, Speech and Signal Processing (ICASSP)}, pages 7669--7673. IEEE, 2020.

\bibitem{Qwen2.5-Omni}
Jin Xu, Zhifang Guo, Jinzheng He, Hangrui Hu, Ting He, Shuai Bai, Keqin Chen, Jialin Wang, Yang Fan, Kai Dang, Bin Zhang, Xiong Wang, Yunfei Chu, and Junyang Lin.
\newblock Qwen2.5-omni technical report.
\newblock {\em arXiv preprint arXiv:2503.20215}, 2025.

\bibitem{voicebench}
Yiming Chen, Xianghu Yue, Chen Zhang, Xiaoxue Gao, Robby~T Tan, and Haizhou Li.
\newblock Voicebench: Benchmarking llm-based voice assistants.
\newblock {\em arXiv preprint arXiv:2410.17196}, 2024.

\bibitem{sdeval}
Junyi Ao, Yuancheng Wang, Xiaohai Tian, Dekun Chen, Jun Zhang, Lu~Lu, Yuxuan Wang, Haizhou Li, and Zhizheng Wu.
\newblock Sd-eval: A benchmark dataset for spoken dialogue understanding beyond words.
\newblock {\em arXiv preprint arXiv:2406.13340}, 2024.

\bibitem{voxdialogue}
Xize Cheng, Ruofan Hu, Xiaoda Yang, Jingyu Lu, Dongjie Fu, Zehan Wang, Shengpeng Ji, Rongjie Huang, Boyang Zhang, Tao Jin, et~al.
\newblock Voxdialogue: Can spoken dialogue systems understand information beyond words?
\newblock In {\em The Thirteenth International Conference on Learning Representations}, 2025.

\bibitem{yang2024air}
Qian Yang, Jin Xu, Wenrui Liu, Yunfei Chu, Ziyue Jiang, Xiaohuan Zhou, Yichong Leng, Yuanjun Lv, Zhou Zhao, Chang Zhou, et~al.
\newblock Air-bench: Benchmarking large audio-language models via generative comprehension.
\newblock {\em arXiv preprint arXiv:2402.07729}, 2024.

\bibitem{chu2024qwen2}
Yunfei Chu, Jin Xu, Qian Yang, Haojie Wei, Xipin Wei, Zhifang Guo, Yichong Leng, Yuanjun Lv, Jinzheng He, Junyang Lin, et~al.
\newblock Qwen2-audio technical report.
\newblock {\em arXiv preprint arXiv:2407.10759}, 2024.

\bibitem{dpo}
Rafael Rafailov, Archit Sharma, Eric Mitchell, Christopher~D Manning, Stefano Ermon, and Chelsea Finn.
\newblock Direct preference optimization: Your language model is secretly a reward model.
\newblock {\em Advances in Neural Information Processing Systems}, 36:53728--53741, 2023.

\bibitem{ppo}
John Schulman, Filip Wolski, Prafulla Dhariwal, Alec Radford, and Oleg Klimov.
\newblock Proximal policy optimization algorithms.
\newblock {\em arXiv preprint arXiv:1707.06347}, 2017.

\bibitem{grpo}
Zhihong Shao, Peiyi Wang, Qihao Zhu, Runxin Xu, Junxiao Song, Xiao Bi, Haowei Zhang, Mingchuan Zhang, YK~Li, Y~Wu, et~al.
\newblock Deepseekmath: Pushing the limits of mathematical reasoning in open language models.
\newblock {\em arXiv preprint arXiv:2402.03300}, 2024.

\bibitem{r1aqa}
Gang Li, Jizhong Liu, Heinrich Dinkel, Yadong Niu, Junbo Zhang, and Jian Luan.
\newblock Reinforcement learning outperforms supervised fine-tuning: A case study on audio question answering.
\newblock {\em arXiv preprint arXiv:2503.11197}, 2025.

\bibitem{cot}
Jason Wei, Xuezhi Wang, Dale Schuurmans, Maarten Bosma, Fei Xia, Ed~Chi, Quoc~V Le, Denny Zhou, et~al.
\newblock Chain-of-thought prompting elicits reasoning in large language models.
\newblock {\em Advances in neural information processing systems}, 35:24824--24837, 2022.

\bibitem{audioreasoner}
Zhifei Xie, Mingbao Lin, Zihang Liu, Pengcheng Wu, Shuicheng Yan, and Chunyan Miao.
\newblock Audio-reasoner: Improving reasoning capability in large audio language models.
\newblock {\em arXiv preprint arXiv:2503.02318}, 2025.

\bibitem{audiocot}
Ziyang Ma, Zhuo Chen, Yuping Wang, Eng~Siong Chng, and Xie Chen.
\newblock Audio-cot: Exploring chain-of-thought reasoning in large audio language model.
\newblock {\em arXiv preprint arXiv:2501.07246}, 2025.

\bibitem{tang2023salmonn}
Changli Tang, Wenyi Yu, Guangzhi Sun, Xianzhao Chen, Tian Tan, Wei Li, Lu~Lu, Zejun Ma, and Chao Zhang.
\newblock Salmonn: Towards generic hearing abilities for large language models.
\newblock {\em arXiv preprint arXiv:2310.13289}, 2023.

\bibitem{du2024cosyvoice}
Zhihao Du, Qian Chen, Shiliang Zhang, Kai Hu, Heng Lu, Yexin Yang, Hangrui Hu, Siqi Zheng, Yue Gu, Ziyang Ma, et~al.
\newblock Cosyvoice: A scalable multilingual zero-shot text-to-speech synthesizer based on supervised semantic tokens.
\newblock {\em arXiv preprint arXiv:2407.05407}, 2024.

\bibitem{cosyvoice2}
Zhihao Du, Yuxuan Wang, Qian Chen, Xian Shi, Xiang Lv, Tianyu Zhao, Zhifu Gao, Yexin Yang, Changfeng Gao, Hui Wang, et~al.
\newblock Cosyvoice 2: Scalable streaming speech synthesis with large language models.
\newblock {\em arXiv preprint arXiv:2412.10117}, 2024.

\bibitem{ji2024mobilespeech}
Shengpeng Ji, Ziyue Jiang, Hanting Wang, Jialong Zuo, and Zhou Zhao.
\newblock Mobilespeech: A fast and high-fidelity framework for mobile zero-shot text-to-speech.
\newblock {\em arXiv preprint arXiv:2402.09378}, 2024.

\bibitem{zhang2023speechgpt}
Dong Zhang, Shimin Li, Xin Zhang, Jun Zhan, Pengyu Wang, Yaqian Zhou, and Xipeng Qiu.
\newblock Speechgpt: Empowering large language models with intrinsic cross-modal conversational abilities.
\newblock {\em arXiv preprint arXiv:2305.11000}, 2023.

\bibitem{llamaomni2}
Qingkai Fang, Yan Zhou, Shoutao Guo, Shaolei Zhang, and Yang Feng.
\newblock Llama-omni2: Llm-based real-time spoken chatbot with autoregressive streaming speech synthesis.
\newblock {\em arXiv preprint arXiv:2505.02625}, 2025.

\bibitem{encodec}
Alexandre D{\'e}fossez, Jade Copet, Gabriel Synnaeve, and Yossi Adi.
\newblock High fidelity neural audio compression.
\newblock {\em arXiv preprint arXiv:2210.13438}, 2022.

\bibitem{ji2024wavtokenizer}
Shengpeng Ji, Ziyue Jiang, Wen Wang, Yifu Chen, Minghui Fang, Jialong Zuo, Qian Yang, Xize Cheng, Zehan Wang, Ruiqi Li, et~al.
\newblock Wavtokenizer: an efficient acoustic discrete codec tokenizer for audio language modeling.
\newblock {\em arXiv preprint arXiv:2408.16532}, 2024.

\bibitem{ji2024language}
Shengpeng Ji, Minghui Fang, Ziyue Jiang, Rongjie Huang, Jialung Zuo, Shulei Wang, and Zhou Zhao.
\newblock Language-codec: Reducing the gaps between discrete codec representation and speech language models.
\newblock {\em arXiv preprint arXiv:2402.12208}, 2024.

\bibitem{zhang2024intrinsicvoice}
Xin Zhang, Xiang Lyu, Zhihao Du, Qian Chen, Dong Zhang, Hangrui Hu, Chaohong Tan, Tianyu Zhao, Yuxuan Wang, Bin Zhang, et~al.
\newblock Intrinsicvoice: Empowering llms with intrinsic real-time voice interaction abilities.
\newblock {\em arXiv preprint arXiv:2410.08035}, 2024.

\bibitem{musicgen}
Jade Copet, Felix Kreuk, Itai Gat, Tal Remez, David Kant, Gabriel Synnaeve, Yossi Adi, and Alexandre D{\'e}fossez.
\newblock Simple and controllable music generation.
\newblock {\em Advances in Neural Information Processing Systems}, 36:47704--47720, 2023.

\bibitem{siuzdak2024snac}
Hubert Siuzdak, Florian Gr{\"o}tschla, and Luca~A Lanzend{\"o}rfer.
\newblock Snac: Multi-scale neural audio codec.
\newblock {\em arXiv preprint arXiv:2410.14411}, 2024.

\bibitem{vita15}
Chaoyou Fu, Haojia Lin, Xiong Wang, Yi-Fan Zhang, Yunhang Shen, Xiaoyu Liu, Haoyu Cao, Zuwei Long, Heting Gao, Ke~Li, et~al.
\newblock Vita-1.5: Towards gpt-4o level real-time vision and speech interaction.
\newblock {\em arXiv preprint arXiv:2501.01957}, 2025.

\bibitem{openomni}
Run Luo, Ting-En Lin, Haonan Zhang, Yuchuan Wu, Xiong Liu, Min Yang, Yongbin Li, Longze Chen, Jiaming Li, Lei Zhang, et~al.
\newblock Openomni: Large language models pivot zero-shot omnimodal alignment across language with real-time self-aware emotional speech synthesis.
\newblock {\em arXiv preprint arXiv:2501.04561}, 2025.

\bibitem{glm4voice}
Aohan Zeng, Zhengxiao Du, Mingdao Liu, Kedong Wang, Shengmin Jiang, Lei Zhao, Yuxiao Dong, and Jie Tang.
\newblock Glm-4-voice: Towards intelligent and human-like end-to-end spoken chatbot.
\newblock {\em arXiv preprint arXiv:2412.02612}, 2024.

\bibitem{kimiaudio}
Ding Ding, Zeqian Ju, Yichong Leng, Songxiang Liu, Tong Liu, Zeyu Shang, Kai Shen, Wei Song, Xu~Tan, Heyi Tang, et~al.
\newblock Kimi-audio technical report.
\newblock {\em arXiv preprint arXiv:2504.18425}, 2025.

\bibitem{wang2024audiobench}
Bin Wang, Xunlong Zou, Geyu Lin, Shuo Sun, Zhuohan Liu, Wenyu Zhang, Zhengyuan Liu, AiTi Aw, and Nancy~F Chen.
\newblock Audiobench: A universal benchmark for audio large language models.
\newblock {\em arXiv preprint arXiv:2406.16020}, 2024.

\bibitem{superb}
Shu-wen Yang, Po-Han Chi, Yung-Sung Chuang, Cheng-I~Jeff Lai, Kushal Lakhotia, Yist~Y Lin, Andy~T Liu, Jiatong Shi, Xuankai Chang, Guan-Ting Lin, et~al.
\newblock Superb: Speech processing universal performance benchmark.
\newblock {\em arXiv preprint arXiv:2105.01051}, 2021.

\bibitem{mmau}
S~Sakshi, Utkarsh Tyagi, Sonal Kumar, Ashish Seth, Ramaneswaran Selvakumar, Oriol Nieto, Ramani Duraiswami, Sreyan Ghosh, and Dinesh Manocha.
\newblock Mmau: A massive multi-task audio understanding and reasoning benchmark.
\newblock {\em arXiv preprint arXiv:2410.19168}, 2024.

\bibitem{spokenwoz}
Shuzheng Si, Wentao Ma, Haoyu Gao, Yuchuan Wu, Ting-En Lin, Yinpei Dai, Hangyu Li, Rui Yan, Fei Huang, and Yongbin Li.
\newblock Spokenwoz: A large-scale speech-text benchmark for spoken task-oriented dialogue agents.
\newblock {\em Advances in Neural Information Processing Systems}, 36:39088--39118, 2023.

\bibitem{chu2023qwen}
Yunfei Chu, Jin Xu, Xiaohuan Zhou, Qian Yang, Shiliang Zhang, Zhijie Yan, Chang Zhou, and Jingren Zhou.
\newblock Qwen-audio: Advancing universal audio understanding via unified large-scale audio-language models.
\newblock {\em arXiv preprint arXiv:2311.07919}, 2023.

\bibitem{rl1}
Lunjun Zhang, Arian Hosseini, Hritik Bansal, Mehran Kazemi, Aviral Kumar, and Rishabh Agarwal.
\newblock Generative verifiers: Reward modeling as next-token prediction.
\newblock {\em arXiv preprint arXiv:2408.15240}, 2024.

\bibitem{deepseekr1}
Daya Guo, Dejian Yang, Haowei Zhang, Junxiao Song, Ruoyu Zhang, Runxin Xu, Qihao Zhu, Shirong Ma, Peiyi Wang, Xiao Bi, et~al.
\newblock Deepseek-r1: Incentivizing reasoning capability in llms via reinforcement learning.
\newblock {\em arXiv preprint arXiv:2501.12948}, 2025.

\bibitem{deepseekv2}
Aixin Liu, Bei Feng, Bin Wang, Bingxuan Wang, Bo~Liu, Chenggang Zhao, Chengqi Dengr, Chong Ruan, Damai Dai, Daya Guo, et~al.
\newblock Deepseek-v2: A strong, economical, and efficient mixture-of-experts language model.
\newblock {\em arXiv preprint arXiv:2405.04434}, 2024.

\bibitem{deepseekv3}
Aixin Liu, Bei Feng, Bing Xue, Bingxuan Wang, Bochao Wu, Chengda Lu, Chenggang Zhao, Chengqi Deng, Chenyu Zhang, Chong Ruan, et~al.
\newblock Deepseek-v3 technical report.
\newblock {\em arXiv preprint arXiv:2412.19437}, 2024.

\bibitem{promptprogramming}
Laria Reynolds and Kyle McDonell.
\newblock Prompt programming for large language models: Beyond the few-shot paradigm.
\newblock In {\em Extended abstracts of the 2021 CHI conference on human factors in computing systems}, pages 1--7, 2021.

\bibitem{agedata}
Khaled Hechmi, Trung~Ngo Trong, Ville Hautam{\"a}ki, and Tomi Kinnunen.
\newblock Voxceleb enrichment for age and gender recognition.
\newblock In {\em 2021 IEEE Automatic Speech Recognition and Understanding Workshop (ASRU)}, pages 687--693. IEEE, 2021.

\bibitem{stepaudio}
Ailin Huang, Boyong Wu, Bruce Wang, Chao Yan, Chen Hu, Chengli Feng, Fei Tian, Feiyu Shen, Jingbei Li, Mingrui Chen, et~al.
\newblock Step-audio: Unified understanding and generation in intelligent speech interaction.
\newblock {\em arXiv preprint arXiv:2502.11946}, 2025.

\bibitem{kim2019audiocaps}
Chris~Dongjoo Kim, Byeongchang Kim, Hyunmin Lee, and Gunhee Kim.
\newblock Audiocaps: Generating captions for audios in the wild.
\newblock In {\em Proceedings of the 2019 Conference of the North American Chapter of the Association for Computational Linguistics: Human Language Technologies, Volume 1 (Long and Short Papers)}, pages 119--132, 2019.

\bibitem{whisperopenai}
Alec Radford, Jong~Wook Kim, Tao Xu, Greg Brockman, Christine McLeavey, and Ilya Sutskever.
\newblock Robust speech recognition via large-scale weak supervision.
\newblock In {\em International conference on machine learning}, pages 28492--28518. PMLR, 2023.

\bibitem{ma2023emotion2vec}
Ziyang Ma, Zhisheng Zheng, Jiaxin Ye, Jinchao Li, Zhifu Gao, Shiliang Zhang, and Xie Chen.
\newblock emotion2vec: Self-supervised pre-training for speech emotion representation.
\newblock {\em arXiv preprint arXiv:2312.15185}, 2023.

\bibitem{audioflamingo2}
Sreyan Ghosh, Zhifeng Kong, Sonal Kumar, S~Sakshi, Jaehyeon Kim, Wei Ping, Rafael Valle, Dinesh Manocha, and Bryan Catanzaro.
\newblock Audio flamingo 2: An audio-language model with long-audio understanding and expert reasoning abilities.
\newblock {\em arXiv preprint arXiv:2503.03983}, 2025.

\bibitem{hu2022lora}
Edward~J Hu, Yelong Shen, Phillip Wallis, Zeyuan Allen-Zhu, Yuanzhi Li, Shean Wang, Lu~Wang, Weizhu Chen, et~al.
\newblock Lora: Low-rank adaptation of large language models.
\newblock {\em ICLR}, 1(2):3, 2022.

\bibitem{swift}
Yuze Zhao, Jintao Huang, Jinghan Hu, Xingjun Wang, Yunlin Mao, Daoze Zhang, Zeyinzi Jiang, Zhikai Wu, Baole Ai, Ang Wang, et~al.
\newblock Swift: a scalable lightweight infrastructure for fine-tuning.
\newblock In {\em Proceedings of the AAAI Conference on Artificial Intelligence}, volume~39, pages 29733--29735, 2025.

\end{thebibliography}

%%%%%%%%%%%%%%%%%%%%%%%%%%%%%%%%%%%%%%%%%%%%%%%%%%%%%%%%%%%%
\newpage

\appendix

\section{Examples in ChatReward-30K}
\label{appendix: example_chatre}
\begin{figure}[htbp]%%图,[htbp]是浮动格式
\centering
\includegraphics[height=3.8cm, width=14cm]{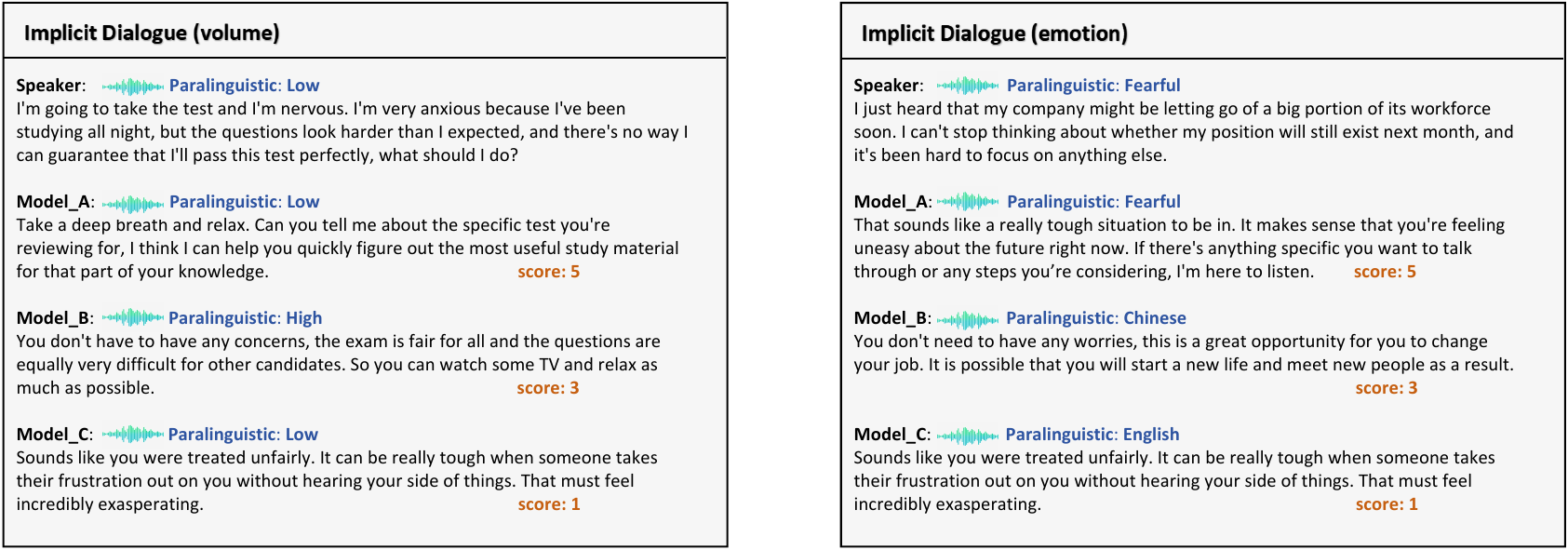}
\caption{The ChatReward-30K dataset encompasses a wide range of both explicit and implicit dialogue scenarios, with responses evaluated by human experts based on model performance.}
\label{figure_chatreward_example3}
\end{figure}

Various examples from the ChatReward-30K dataset are illustrated in Figure~\ref{figure_chatreward_example3}.

\begin{figure}[htbp]%%图,[htbp]是浮动格式
\centering
\includegraphics[height=8cm, width=14cm]{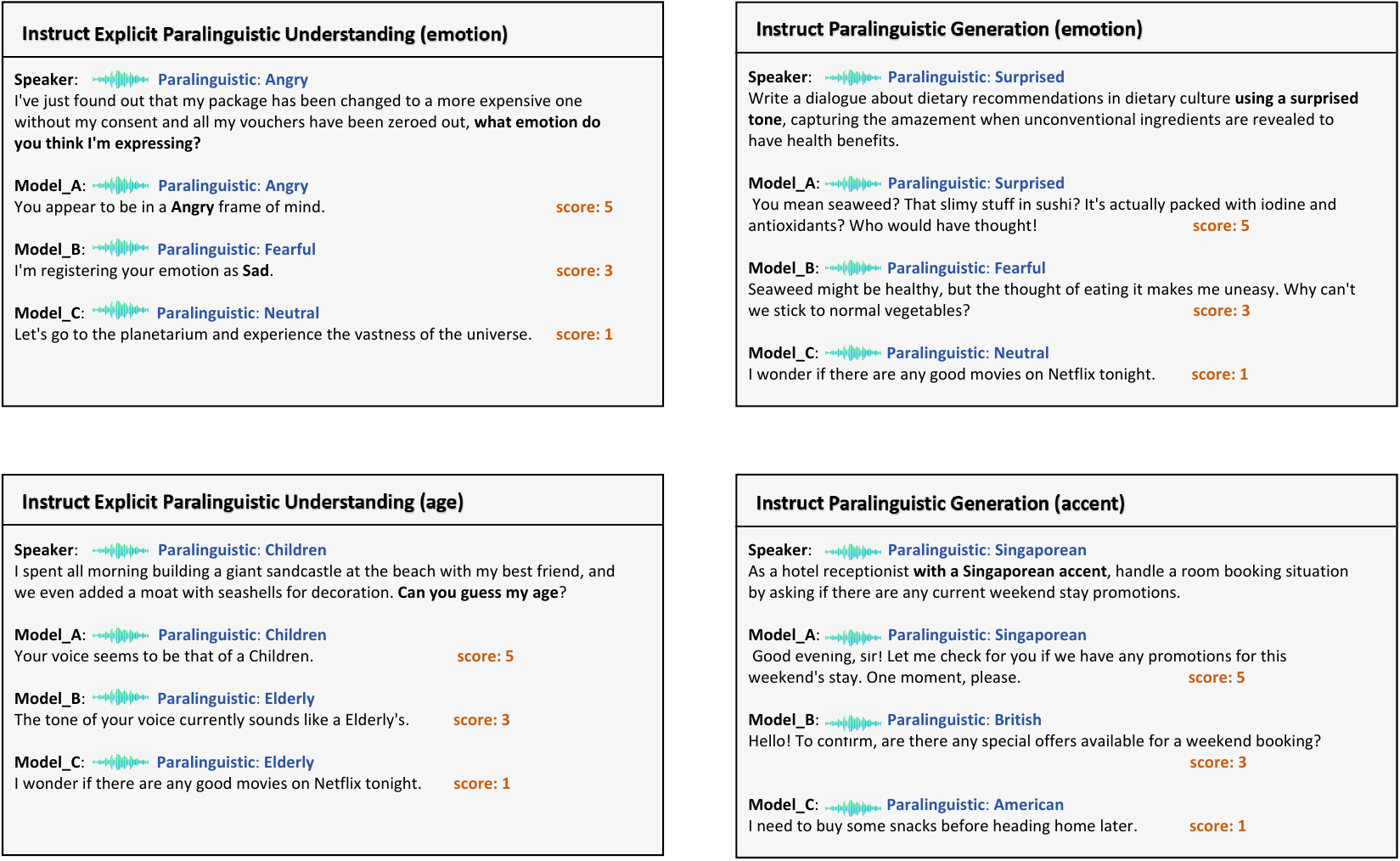}
% \caption{WavReward can be applied to evaluate various dialogue scenarios, including both explicit instruction and implicit dialogues. It directly accepts speech-to-speech dialogue as input, evaluating the conversational coherence at both the textual and acoustic levels, and providing the final score.}
\label{figure_chatreward_example1}
\end{figure}

\begin{figure}[htbp]%%图,[htbp]是浮动格式
\centering
\includegraphics[height=4.2cm, width=14cm]{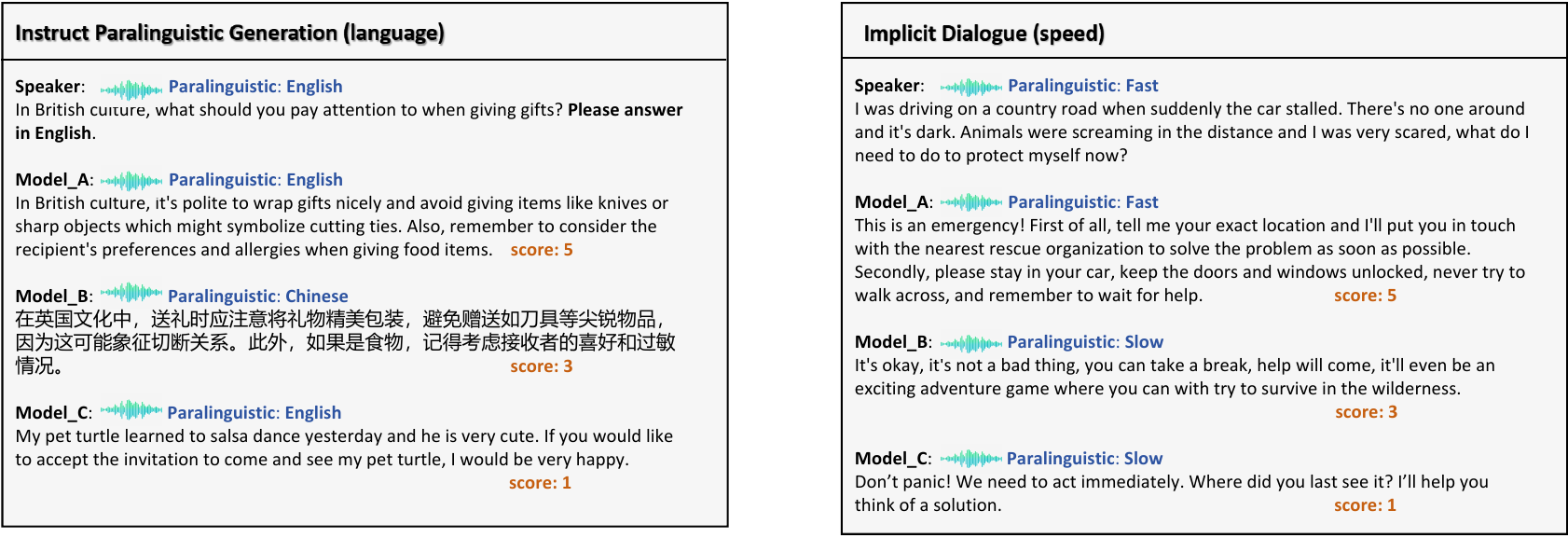}
% \caption{WavReward can be applied to evaluate various dialogue scenarios, including both explicit instruction and implicit dialogues. It directly accepts speech-to-speech dialogue as input, evaluating the conversational coherence at both the textual and acoustic levels, and providing the final score.}
\label{figure_chatreward_example2}
\end{figure}

\section{Acoustic information in ChatReward-30K}
\label{appendix: acoustic_in}

The specific categories, sample quantities, and durations of all acoustic information in ChatReward-30K are detailed in Table~\ref{tab:acoustic_information}.

\begin{table}[htbp]
    \centering
    \small
   \caption{Detailed statistics of the corresponding subsets of each attribute in ChatReward-30K.}
   \begin{adjustbox}{width=\textwidth}
    \begin{tabular}{c|lcc}
      \toprule
      Attributes & Categories & Samples & Duration \\
      \midrule
      Gender & male, female & 2177  & 9.56Hours  \\
      Age & children, elderly, middle-aged, adolescent & 2070 & 8.36Hours  \\
      Language & chinese, english & 3583  & 16.23Hours  \\
      Accent & indian, canadian, british, singaporean, american, australian & 1618 & 5.70Hours  \\
      \midrule
      Emotion & neutral, happy, sad, angry, surprised, disgusted, fearful & 9470 & 52.04Hours  \\
      Pitch & low, high, normal & 853 &  3.65Hours \\
      Speed & slow, normal, fast & 2303 &  10.95Hours \\
      Volume & low, normal, high & 2054 &  7.53Hours \\
      \midrule
      \multirow{2}{*}{Audio} & laughing, crying,  bee, bird, car, cat, chirping, clapping,
coughing, dog, screaming& \multirow{2}{*}{4081}  & \multirow{2}{*}{15.38Hours}  \\
      & duck, horse, ice, knocking, ocean, pig, police, sneezing, thunder, waterfall burbling& & \\
      % \multirow{2}{*}{Music} & violin, cello, double bass, guitar, acoustic guitar, electric guitar, banjo,  harp, ukulele &  &   \\
      % &oboe, clarinet, bassoon, saxophone, trumpet, trombone, piano, mandolin, flute& & \\
      \midrule
      \textbf{Overall} &  & \textbf{28209} & \textbf{129.40Hours}  \\
      \bottomrule
    \end{tabular}
    \end{adjustbox}
    \label{tab:acoustic_information}
\end{table}

\section{Prompt Programming Template for ChatReward-30K}
\label{appendix: prompt}
The sample prompt template for ChatReward-30K is illustrated in the Figure~\ref{figure_prompt1} below:

\begin{figure}[htbp]%%图,[htbp]是浮动格式
\centering
\includegraphics[height=8cm, width=14cm]{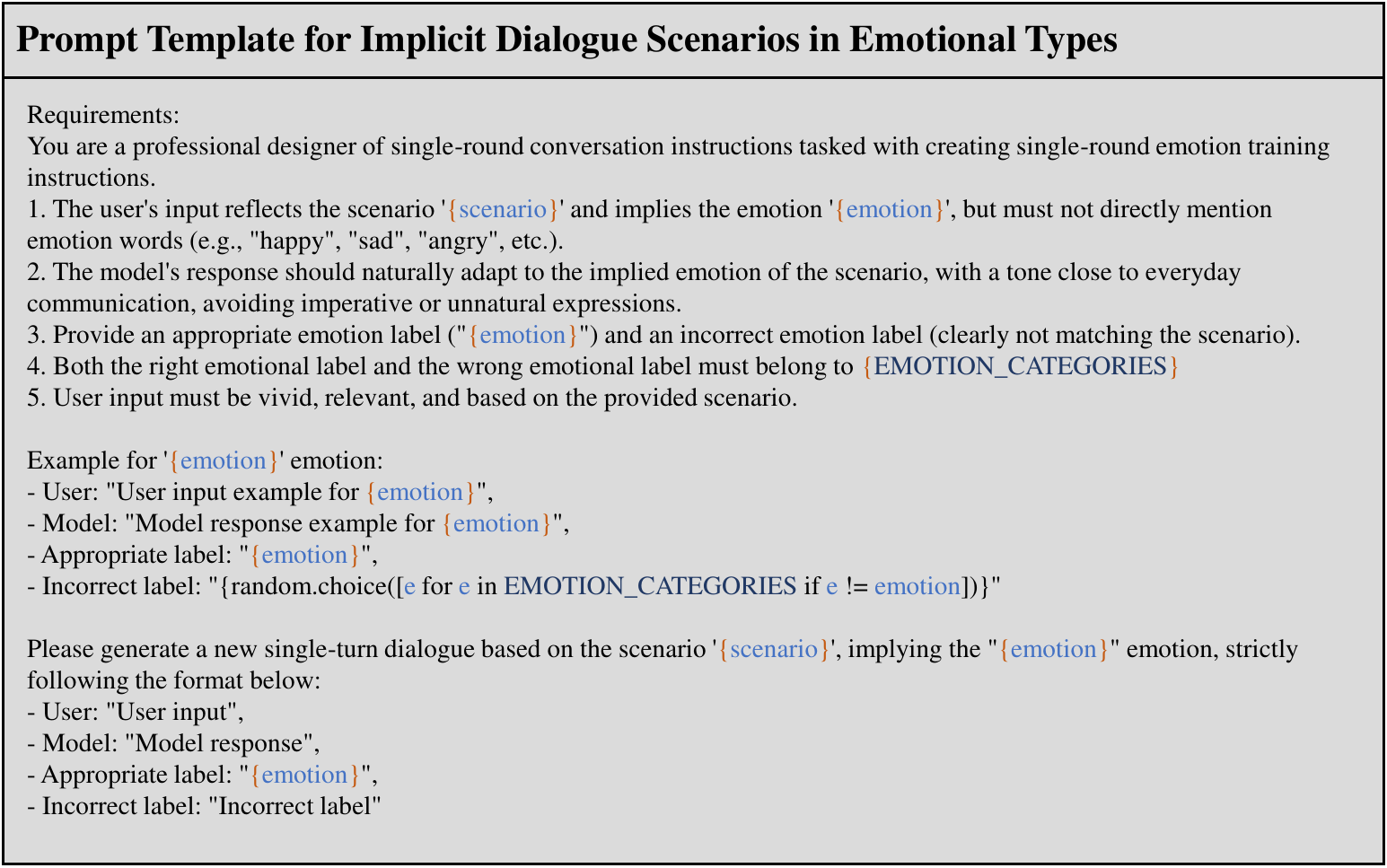}
\caption{The emotion prompt template for ChatReward-30K.}
\label{figure_prompt1}
\end{figure}

\section{Prompt Programming Template (Without COT) for WavReward}
\label{appendix: prompt_think}

\begin{figure}[htbp]%%图,[htbp]是浮动格式
\centering
\includegraphics[height=8cm, width=14cm]{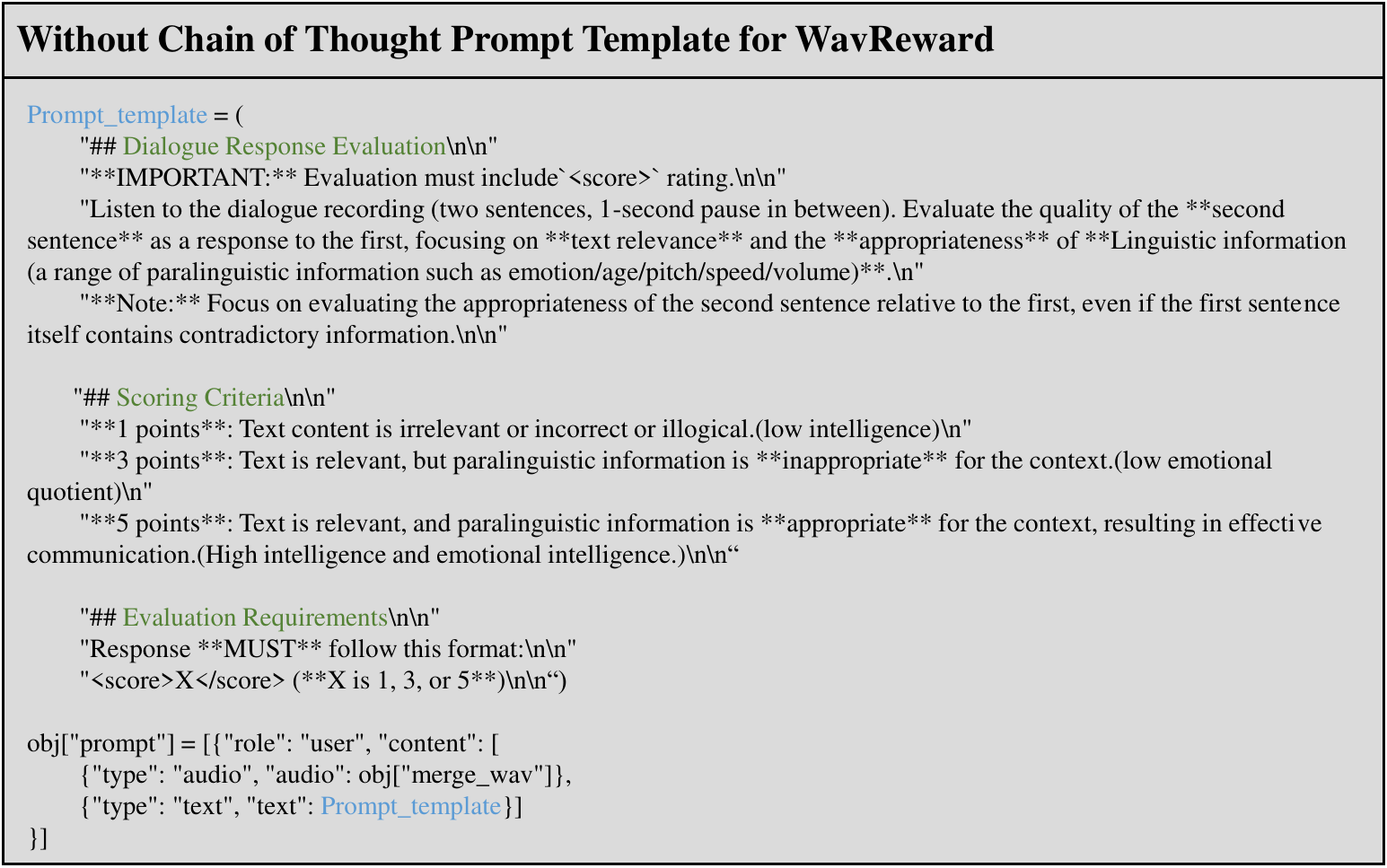}
\caption{The ablation prompt template for WavReward.}
\label{figure_prompt2}
\end{figure}

The ablation text prompt template for WavReward w/o COT is shown in Figure~\ref{figure_prompt2}.

\section{Statistics of ChatReward-30K}
\label{appendix_statistics}
\begin{figure}[htbp]%%图,[htbp]是浮动格式
\centering
\includegraphics[height=6cm, width=12cm]{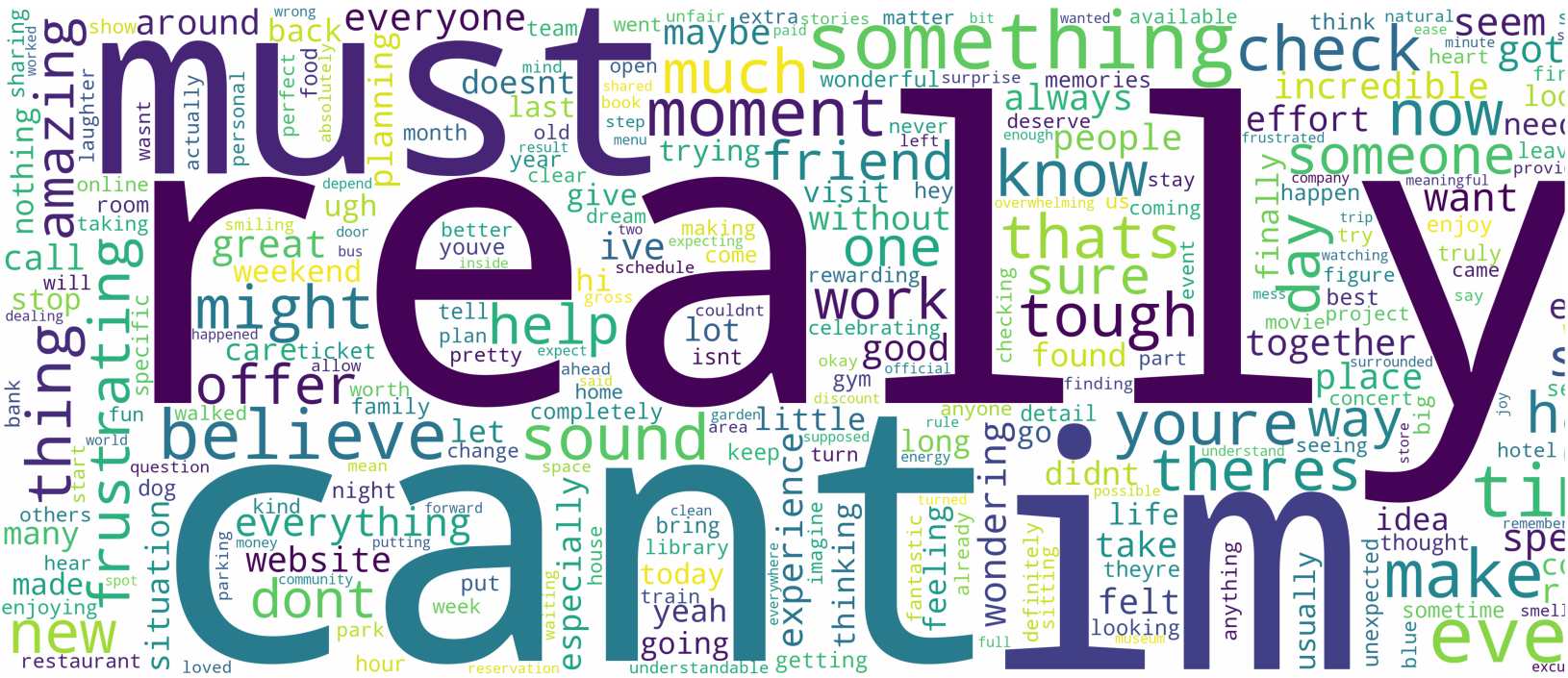}
\caption{Word Cloud of ChatReward-30K.}
\label{figure_chatreward_data3}
\end{figure}

\begin{figure}[htbp]%%图,[htbp]是浮动格式
\centering
\includegraphics[height=8cm, width=9.5cm]{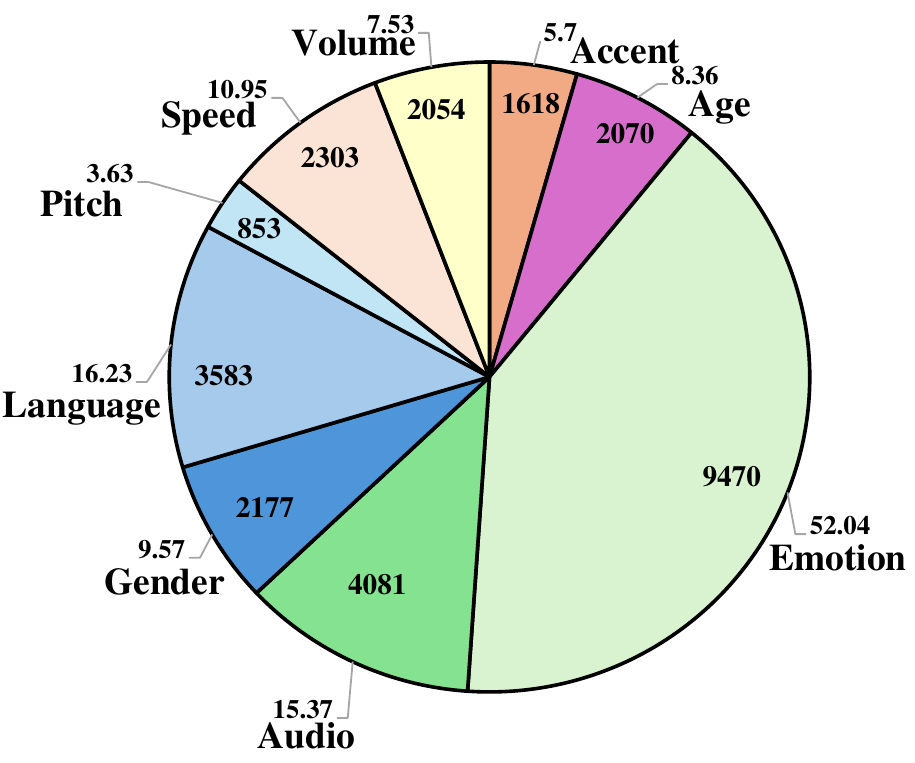}
\caption{Statistics of different acoustic attribute in ChatReward-30K.}
\label{figure_chatreward_data2}
\end{figure}

tatistics of different acoustic attribute in ChatReward-30K is illustrated in the Figure~\ref{figure_chatreward_data2}. Word Cloud of ChatReward-30K is shown in Figure~\ref{figure_chatreward_data3}.

% \begin{table}[htbp]
%     \centering
%     \small
%    \caption{Comparison of recentl.}
%    \begin{adjustbox}{width=\textwidth}
%     \begin{tabular}{c|cccc}
%       \toprule
%       Model & Single Codebook & Speech & Music\&Sound & Separate/Unified model \\
%       \midrule
%       \textbf{UniCodec} & $\usym{2714}$ & $\usym{2714}$ & $\usym{2714}$ & \textbf{Unified} \\
%       \bottomrule
%     \end{tabular}
%     \end{adjustbox}
%     \label{tab:acoustic_information}
% \end{table}

%%%%%%%%%%%%%%%%%%%%%%%%%%%%%%%%%%%%%%%%%%%%%%%%%%%%%%%%%%%%

\end{document}